\documentclass[prb,twocolumn,amsmath,amssymb,superscriptaddress,showpacs,floatfix]{revtex4-2}

\usepackage[english]{babel}
\usepackage{amsfonts}
\usepackage{graphicx}
\usepackage{bm}
\usepackage{dcolumn}
\usepackage{booktabs}
\usepackage{xcolor}
\usepackage[colorlinks=true,linkcolor=blue,citecolor=blue,urlcolor=blue]{hyperref}

\newcommand{\bk}{\mathbf{k}}
\newcommand{\bq}{\mathbf{q}}
\newcommand{\bQ}{\mathbf{Q}}
\newcommand{\bzavg}[1]{\bigl\langle #1 \bigr\rangle}
\newcommand{\nB}{n_{B}}
\newcommand{\Hhat}{\hat{H}}
\newcommand{\Htheta}{\hat{H}_{\theta}}
\newcommand{\tb}{\tilde{b}}
\newcommand{\tq}{\tilde{q}}
\newcommand{\tm}{\tilde{m}}
\newcommand{\ket}[1]{\lvert #1\rangle}
\newcommand{\bra}[1]{\langle #1\rvert}
\newcommand{\braket}[2]{\langle #1\vert #2\rangle}

\begin{document}

\preprint{APS/123-QED}

\title{Exact high-order spectral moments and moment-based reconstruction of polaron spectra at finite temperature for arbitrary linear electron–phonon coupling}

\author{Jeet Shannigrahi}
\affiliation{Department of Physics and Astronomy, University of British Columbia, Vancouver, British Columbia, Canada, V6T 1Z1}
\affiliation{Stewart Blusson Quantum Matter Institute, University of British Columbia, Vancouver, British Columbia, Canada, V6T 1Z4}

\author{Mona Berciu}
\affiliation{Department of Physics and Astronomy, University of British Columbia, Vancouver, British Columbia, Canada, V6T 1Z1}
\affiliation{Stewart Blusson Quantum Matter Institute, University of British Columbia, Vancouver, British Columbia, Canada, V6T 1Z4}

\date{\today}

\begin{abstract}
We present a family of exact and efficient recurrence relations for the frequency moments of the single-electron addition spectral function of a generic single-band electron--phonon Hamiltonian with linear coupling to harmonic phonons. Working in the limit of zero carrier density, the formulation applies at zero and finite temperature and generalizes to arbitrary dimensions and lattice geometry. Using the Holstein and Peierls models coupled to Einstein and dispersive phonons as examples, we obtain closed-form expressions for moments of far higher order than those available in the literature, at modest computational cost. Beyond providing stringent benchmarks for numerical methods and experimental spectra, the high-order moments serve as input to approximate reconstructions of polaron spectra at weak-to-intermediate coupling and low temperature. The resulting spectral functions are exact in the weak-coupling and atomic limits, and respect rigorous bounds on the polaron energy and weight that we show the self-consistent Born approximation violates throughout the coupling--adiabaticity plane. The approach therefore offers an inexpensive yet accurate approximation, especially for unexplored or numerically intractable models, before computationally demanding exact methods are applied.
\end{abstract}

\maketitle

\section{Introduction}
\label{sec:intro}
The single-particle spectral function $A(\omega)$ is one of the first quantities one determines in condensed matter physics when analysing the behaviour of complex, interacting systems. This is because it gives critical insights into the eigen-spectrum of the system,  acting as a density operator for the eigenstates. $A(\omega)$ also holds experimental relevance, and can be directly measured experimentally through Angle Resolved (Inverse) Photo-Emission Spectroscopy (ARIPES)~\cite{Damascelli2003}. Consequently, a whole gamut of theoretical and computational approaches have been developed to compute the spectral function (and its associated Green's function) reliably for interacting systems. Concurrently,  the frequency moments
\begin{equation}
M_n(\bk)=\int_{-\infty}^{\infty}d\omega\,\omega^{n}A(\bk,\omega)
\label{eq:Mn_def}
\end{equation}
form very useful exact identities derived from the spectral function. They are integral constraints---``sum rules''---that a spectral function must satisfy, irrespective of its detailed shape. These identities  validate calculations for real systems and for a theoretical/computational method to be deemed physically useful, the sum rules of the computed spectral function must be satisfied up to an arbitrary order. 
Within condensed matter physics and specifically in problems of strongly correlated systems,  moments have seen broad application as diagnosis of energy and spectral weight.  Harris and Lange used moments of the individual Hubbard bands to characterize their energies at strong coupling~\cite{HarrisLange1967}; Nolting fixed the weights and peak positions of a two-pole spectral-density ansatz by equating moments for a simple-cubic lattice \cite{Nolting1972}; White used the exact first and second moment identities as constraints (with some tolerance) in the maximum-entropy continuation of quantum Monte Carlo data for the two-dimensional Hubbard model~\cite{White1991}. More recently and of specific interest to the polaron community, there is a long thread of work done by Freericks and collaborators which derived exact low-order moment identities for the local retarded Green's functions and self-energies provide one of the few unbiased benchmarks for driven systems~\cite{TurkowskiFreericks2006,TurkowskiFreericks2008,FreericksTurkowski2009,Freericks2013}. In the large-$U$ Holstein--Hubbard model, R\"osch and collaboratorss used self-energy sum rules to show that vertex corrections are essential for an electron self-energy sum rule in the undoped system, while a phonon self-energy sum rule in a doped system can be satisfied without them~\cite{Rosch2007}. 

Another direction in which moments have been considered useful is the prospect of reconstructing the  spectrum.  Early examples include using moments to reconstruct densities of states in tight-binding bands \cite{CyrotLackmann1967,GaspardCyrotLackmann1973}.  For a positive spectral measure, orthogonalizing the powers of frequency gives a three-term recurrence and a continued-fraction representation of its Green's function $G(\bk,z)$~\cite{HaydockHeineKelly1972,Haydock1980,ViswanathMuller1994,Simon1998}. An $N$-dimensional Jacobi block fixed by the moments yields a positive Gaussian quadrature reproducing $M_0,\ldots,M_{2N-1}$ exactly ~\cite{GolubWelsch1969}. Truncating the resultant continued fraction after $N$ exact levels and closing it with a terminator beyond that exact block preserves this set of moments and  produces a spectral function that is positive by construction  ~\cite{HaydockNex1985}. Recent work has analyzed the error from stitching such tails under specified assumptions~\cite{Pinna2025} and proposed a self-consistent moment quadrature~\cite{Kruchinin2026}. Related Nevanlinna and moment-problem methods use Euclidean correlator data to bound smeared spectra~\cite{Abbott2026}. Other techniques for reconstruction include the kernel polynomial method through the use of modified Chebyshev moments ~\cite{Weisse2006}, and maximum entropy which selects a distribution subject to a finite number of moment constraints~\cite{MeadPapanicolaou1984}.In practice, the resolution of the spectral function reconstructed  from a finite sequence depends critically on the number and precision of the usable moments ~\cite{Weisse2006,Gautschi2004}.

Electron--phonon problems represent an interesting domain of application. Within the field, arguably the earliest effort to relate moments to relevant physical quantities was Kornilovitch's work\cite{Kornilovitch2002} which showed the possibility of accessing information about the  early stages of polaron formation through quantities constructed out of low-lying moments. Kornilovitch derived the one-polaron sum rules as nested commutators of the charge carrier operator with the Hamiltonian and  formulated them for the symmetrized spectral function at finite carrier density (they reduce to the electron-addition moments in the low-density/insulator limit used in our work). He evaluated the commutator explicitly for the first two moments of a generic single-band polaron Hamiltonian coupled to optical phonons and the first four of the Holstein model, and pointed out that the bare dispersion and the coupling constant can be read off from $M_1$ and $M_2$ of a momentum-resolved spectrum~\cite{Kornilovitch2002}. 

Berciu and collaborators established that for an approximation/technique to be accurate at all couplings , it is important that not only does it meet the first few sum rules exactly but also satisfy \emph{all} higher ones with good accuracy ~\cite{Berciu2006,Goodvin2006}. It singled out the Born approximation (also referred to as the Migdal approximation) and the self-consistent Born approximation (SCBA) for matching the first four moments exactly but substantially deviating at higher orders. It is easy to see why through diagrammatics, BA drops all terms beyond the first order diagram and SCBA drop all crossed terms.  Consequently, BA and SCBA are limited to a small region of the parameter space, to the weak coupling regime where contributions from higher orders of diagrams (intuitively, more phonons) are relatively reduced.

Mitri\'c and coworkers established that the second-order cumulant expansion (CE) reproduces the Holstein moments exactly through $M_4$, and used the discrepancy at $M_5$ to delimit the range of validity of the CE~\cite{Mitric2022,Mitric2023,MitricThesis2023}. They also recently derived the general finite-temperature sum rules $M_0$--$M_4$ for an arbitrary momentum-dependent vertex $g_{\bk,\bq}$ and used them to benchmark the CE for the Peierls/SSH model~\cite{Mitric2026}. Nonequilibrium versions of the lowest Holstein sum rules exist as well~\cite{Freericks2014}.
 
Moments have seen extensive use as litmus tests for numerically exact methods: finite-T versions of $M_1$--$M_3$ at finite temperature were used to validate the variational exact-diagonalization finite-temperature Lanczos (VED--FTLM) spectra of Bon\v{c}a, Trugman, and Berciu~\cite{Bonca2019}; the electron-addition $M_2$ was checked against finite-temperature density-matrix renormalization group (DMRG) data~\cite{Meisner2020}; However,  to the best of our knowledge, there never has been a concerted effort to use sum moments beyond trivial checks on numerical data or reconstructing spectral functions. The latter can be immediately attributed to the difficulty of obtaining exact moments of high order. This work removes the latter constraint and seeks to show a new perspective on utilising moments.

In the following paragraphs, we briefly summarize the past formulations for the moments of the single particle spectral function in the polaron problem and the exact bottleneck we solve. Writing $\omega^{n}=\bigl(i\,d/d\tau\bigr)^{n}e^{-i\omega\tau}\big|_{\tau=0}$ under the integral in Eq.~\eqref{eq:Mn_def} converts the moment into the $n$-th time derivative of the retarded Green's function at $\tau\to0^{+}$,
\begin{equation}
M_n(\bk)=
-\frac{1}{\pi}\,\mathrm{Im}\,\Bigl(i\frac{d}{d\tau}\Bigr)^{n}
\int d\omega\,e^{-i\omega\tau}G(\bk,\omega)\Big|_{\tau\to0^{+}} .
\label{eq:Mn_tau}
\end{equation}
At $T=0$ and in the absence of other carriers, $G(\bk,\tau)=-i\Theta(\tau)\bra{0}c_{\bk}e^{-iH\tau}c^{\dagger}_{\bk}\ket{0}$ with $H\ket{0}=0$,  and one obtains the compact form used in Ref.~\cite{Goodvin2006},
\begin{equation}
M_n(\bk)=\bra{0}c_{\bk}\,H^{n}\,c^{\dagger}_{\bk}\ket{0}.
\label{eq:Mn_T0_intro}
\end{equation}
At finite temperature the phonon bath is thermally populated and the trace runs over configurations $\{n\}$ with energies $E_{\{n\}}$. Pushing the derivative inside the trace and using the Heisenberg equation of motion $dc_{\bk}/dt=-i[c_{\bk},H]$ gives instead the nested-commutator form of Refs.~\cite{Kornilovitch2002,MitricThesis2023},
\begin{equation}
M_n(\bk)=
\Bigl\langle
\underbrace{[\cdots[[c_{\bk},H],H]\cdots,H]}_{n\ \text{times}}\;c^{\dagger}_{\bk}
\Bigr\rangle_{T}.
\label{eq:Mn_comm_intro}
\end{equation}
One may equally write Eq.~\eqref{eq:Mn_comm_intro} as $Z_{0}^{-1}\sum_{\{n\}}e^{-\beta E_{\{n\}}}\bra{n}c_{\bk}(H-E_{\{n\}})^{n}c^{\dagger}_{\bk}\ket{n}$, i.e.\ as a thermal average of powers of $H$ measured from the energy of the initial bath configuration. While this is more transparent---the excitation energy is counted from the state the electron was injected into---but it is no easier to evaluate: the configuration dependence of $E_{\{n\}}$ prevents the collapse to Eq.~\eqref{eq:Mn_T0_intro}, and one is left with a binomial expansion of $(H-E_{\{n\}})^{n}$ sampled over every bath configuration. 

In practice the nested commutator is expanded by hand or by computer algebra, and the resulting momentum sums are evaluated one by one. While clever manipulations exist such as for the $T = 0$ case, for even $n=2p$ one may write $M_{2p}=\langle H^{p}k|H^{p}k\rangle=\sum_m\|(H^{p}\ket{k})_m\|^{2}$, and for odd $n$, $M_{2p+1}=\langle H^{p}k|H^{p+1}k\rangle$ and solve the two separately, exploiting the orthogonality of the final states,  the number of terms grows  rapidly with $n$, resulting in a very cumbersome process. This is why the general-vertex results stop at $M_4$~\cite{Mitric2026}, the Holstein results at $M_6$~\cite{Goodvin2006}, $M_8-M_9$~\cite{MitricThesis2023, Mitric2026}).

On the other hand, expanding $G(\bk,z)=\int d\nu\,A(\bk,\nu)/(z-\nu)$ at large $|z|$ identifies $M_n$ with the coefficient of $z^{-(n+1)}$ in the high-frequency tail of the Green's function. Since a self-energy diagram with $\ell$ phonon lines contains $2\ell+1$ bare propagators, it first contributes at order $z^{-(2\ell+1)}$, so $M_n$ receives contributions only from diagrams with $\ell\le\lfloor n/2\rfloor$ phonon lines~\cite{Goodvin2006}. Every sum rule is a finite polynomial in the Hamiltonian parameters, and an approximation that keeps all diagrams with up to $\ell$ lines exactly is guaranteed to be exact through $M_{2\ell+1}$---the origin of the Born and SCBA deviations mentioned earlier. Second, the moments are the only closed-form, non-perturbative information about $A(\bk,\omega)$ that is valid at all couplings in the thermodynamic limit for essentially any model in the class considered here, which makes them an excellent starting point for a computationally cheap first-order assessment of models for which no exact numerical solution exists.

In this paper we showcase a family of exact recurrence relations to evaluate Eq.~\eqref{eq:Mn_comm_intro}. Rather than expanding the nested commutator, we propagate the state $c^{\dagger}_{\bk}\ket{0}$ in a phonon-number-resolved basis and derive a closed recurrence for the amplitudes of its $m$-phonon components; the moment is read off from the zero-phonon amplitude after $n$ steps. The recurrence has three terms at $T=0$ and five at $T>0$, it is exact, and it is agnostic to the form of the vertex $g(\bk,\bq)$ and of the phonon dispersion $\Omega(\bq)$. The finite-temperature version is obtained by thermofield doubling: the thermal trace becomes a vacuum expectation value in a doubled phonon space, and the $T=0$ construction applies verbatim with two boson species. It delivers closed-form moments to $M_{28}$ and beyond for the Holstein, Peierls/SSH, and dispersive-phonon Holstein models at arbitrary temperature. We then use these high-order moments as input to continued-fraction reconstructions of $A(\bk,\omega)$ and show, both through the sum rules it satisfies and through rigorous bounds on the polaron energy and weight, that the result is more accurate than Born-type approximations and becomes exact in the weak-coupling and atomic asymptotic limits.

The paper is organized as follows. Section~\ref{sec:recurrence} sets up the model, derives the zero-temperature recurrence in a phonon-number-resolved basis, extends it to finite temperature by thermofield doubling, and discusses its reach and the class of models for which the closed-form expressions of moments exist. Section~\ref{sec:validation} uses the high-order moments to establish, order by order, which sum rules the standard analytical approximations---momentum average, dynamical mean-field theory, Born-type self-energies, ---satisfy, and which diagram is responsible for the first failure in each case. Section~\ref{sec:reconstruction} turns the moments into spectral functions: it describes how the continued fraction is constructed and the limitations of the formulation. Section~\ref{sec:applications} derives rigorous upper bounds on the polaron energy and quasiparticle weight at $T=0$ and uses them to map, across the coupling--adiabaticity plane, where the Born and MA approximations exceed the bounds for the SSH and Holstein models. Section~\ref{sec:conclusions} summarizes and points to future extensions.

\section{Exact recurrence relations for the spectral moments}
\label{sec:recurrence}

\subsection{Model, conventions, and scope}
\label{sec:model}

We consider a single electron coupled linearly to a harmonic lattice in a one-band model,
\begin{align}
H={}&\sum_{\bk}\varepsilon_{\bk}\,c^{\dagger}_{\bk}c^{\phantom{\dagger}}_{\bk}
+\sum_{\bq}\Omega(\bq)\,b^{\dagger}_{\bq}b^{\phantom{\dagger}}_{\bq}
\nonumber\\
&+\frac{1}{\sqrt{N}}\sum_{\bk,\bq}
g(\bk,\bq)\,
c^{\dagger}_{\bk-\bq}c^{\phantom{\dagger}}_{\bk}
\bigl(b^{\dagger}_{\bq}+b^{\phantom{\dagger}}_{-\bq}\bigr),
\label{eq:H}
\end{align}
where $N$ is the number of unit cells, $c^{\dagger}_{\bk}$ creates the electron with bare dispersion $\varepsilon_{\bk}$, and $b^{\dagger}_{\bq}$ creates a phonon of energy $\Omega(\bq)$. Let us note the following about this specific Hamiltonian:

\emph{(i) Vertex convention and Hermiticity.} The first slot of $g(\bk,\bq)$ carries the \emph{initial} electron momentum and the second the momentum of the \emph{created} phonon, so the electron scatters $\bk\to\bk-\bq$. Writing the general linear coupling as $N^{-1/2}\sum_{\bk\bq}[g_1 c^{\dagger}_{\bk-\bq}c_{\bk}b^{\dagger}_{\bq}+\mathrm{h.c.}]$ and collecting terms shows that the compact form~\eqref{eq:H}, in which the same $g$ multiplies $b^{\dagger}_{\bq}$ and $b_{-\bq}$, holds if and only if
\begin{equation}
g(\bk,\bq)=g^{*}(\bk-\bq,-\bq).
\label{eq:hermiticity}
\end{equation}
 In the convention of Refs.~\cite{MarchandThesis2011,Mitric2026}, where the interaction is written $c^{\dagger}_{\bk+\bQ}c_{\bk}(b^{\dagger}_{-\bQ}+b_{\bQ})$ and the second slot holds the momentum transfer $\bQ=\bk_{\rm f}-\bk_{\rm i}$, one has $g(\bk,\bq)=g_{\rm M}(\bk,-\bq)$ and Eq.~\eqref{eq:hermiticity} reads $g_{\rm M}(\bk,\bQ)=g^{*}_{\rm M}(\bk+\bQ,-\bQ)$. While either convention is suitable, it is important that a chosen convention is maintained throughout.

\emph{(ii) Phonon frequency conventions.} We take $\Omega(\bq)=\Omega(-\bq)>0$ and omit the zero-point constant, so that the phonon vacuum satisfies $H_{\rm ph}\ket{0}=0$. Evenness of $\Omega(\bq)$ ensures inversion or time-reversal symmetry, so that $+\bq$ and $-\bq$ phonons are thermally populated identically.

\emph{(iii) Scope.} Equation~\eqref{eq:H} is the most general \emph{one-band, single-branch, linear} electron--phonon Hamiltonian. It excludes multi-band and multi-branch couplings, terms of second and higher order in the lattice displacement (quadratic), and anharmonicity, all of which would break the two-term $(b^{\dagger}+b)$ structure that the recurrence exploits. Within the class~\eqref{eq:H}, however, every model of interest---Holstein~\cite{Holstein1959}, Peierls/Su--Schrieffer--Heeger (SSH)~\cite{SSH,Barisic1970,Marchand2010}, bond-Peierls~\cite{Carbone2021bond}, breathing-mode~\cite{Lau2007}, lattice Fr\"ohlich~\cite{AlexandrovKornilovitch1999} - differs from the others {only} through the pair $\{\varepsilon_{\bk},g(\bk,\bq)\}$ and the dispersion $\Omega(\bq)$. A single recurrence with a customizable vertex therefore covers all of them. Extensions of Eq.~\eqref{eq:H} to different lattice topologies (triangular, etc), including next nearest neighbours or to $d>1$ require nothing more than the replacement of $\epsilon_k$ and its sums.

In the thermodynamic limit we use
\begin{equation}
\frac{1}{N}\sum_{\bq}\;\longrightarrow\;
\bzavg{\,\cdot\,}_{\bq}
\equiv\frac{1}{(2\pi)^{d}}\int_{\rm BZ}\! d^{d}q ,
\label{eq:bzavg}
\end{equation}
and specialize to $d=1$ with $\varepsilon_k=-2t\cos k$ where explicit results are quoted. We set $\hbar=k_B=1$ and represent temperature in units of the phonon frequency ($\Omega$). 

\subsection{Moments as nested commutators}
\label{sec:moments}

The object of interest is the electron-addition spectral function of a single carrier injected into a phonon bath in thermal equilibrium. In the zero-electron sector $H=H_{\rm ph}$, so the exact initial state is $\rho=\ket{0_{\rm el}}\bra{0_{\rm el}}\otimes e^{-\beta H_{\rm ph}}/Z_{\rm ph}$. With $H_{\rm ph}\ket{\alpha}=E_\alpha\ket{\alpha}$ and $p_\alpha=e^{-\beta E_\alpha}/Z_{\rm ph}$,
\begin{equation}
A(k,\omega)=\sum_{\alpha}p_\alpha\sum_{f}\bigl|\bra{f}c^{\dagger}_{k}\ket{\alpha}\bigr|^{2}\,\delta(\omega-E_f+E_\alpha),
\label{eq:Akw}
\end{equation}
and the moments of Eq.~\eqref{eq:Mn_def} become
\begin{align}
M_n(k,T)
&=\sum_{\alpha}p_\alpha\,\bra{\alpha}c_{k}\,(H-E_\alpha)^{n}\,c^{\dagger}_{k}\ket{\alpha}
\label{eq:M_shifted}\\
&=\Bigl\langle
\underbrace{[\cdots[[c_{\bk},H],H]\cdots,H]}_{n\ \text{times}}\;c^{\dagger}_{\bk}
\Bigr\rangle_{T}
\label{eq:M_comm}
\end{align}

which is Eq.~\eqref{eq:Mn_comm_intro}. The subtraction $H\to H-E_\alpha$ in Eq.~\eqref{eq:M_shifted} is essential: the excitation energy is measured from the bath energy of the configuration the electron was added to, and $M_n\neq\mathrm{Tr}[\rho\,c_kH^nc^{\dagger}_k]$ already at $n=1$, where the two differ by the extensive quantity $\langle H_{\rm ph}\rangle_T$. At $T=0$ only $\ket{\alpha}=\ket{0}$ survives, $E_0=0$ by convention (ii), and
\begin{equation}
M_n(k)=\bra{0}c_{k}H^{n}c^{\dagger}_{k}\ket{0}=\bra{k}H^{n}\ket{k},
\qquad \ket{k}\equiv c^{\dagger}_{k}\ket{0}.
\label{eq:M_T0}
\end{equation}
The first two moments follow immediately from Eq.~\eqref{eq:M_comm} for any vertex: $M_0=1$ and $M_1=\varepsilon_k$.

\subsection{Phonon-number-resolved amplitudes at $T=0$}
\label{sec:basis}

Equation~\eqref{eq:M_T0} enables us to interpret $M_n$ as the overlap of $H^{n}\ket{k}$ with $\ket{k}$. The idea of the recurrence is to track the full state $H^{n}\ket{k}$, sector by sector in phonon number, rather than to expand the operator $H^{n}$. Since $H$ conserves the total momentum $K=k_{\rm el}+\sum_i q_i$ and $\ket{k}$ has $K=k$, every state generated by $H^{n}\ket{k}$ lies in the one-electron, fixed-$K$ sector and is a linear combination of
\begin{equation}
\ket{k;q_1,\dots,q_m}
\equiv c^{\dagger}_{\kappa_m}\,b^{\dagger}_{q_1}\cdots b^{\dagger}_{q_m}\ket{0},
\qquad
\kappa_m\equiv k-\sum_{i=1}^{m}q_i ,
\label{eq:basis}
\end{equation}
i.e.\ one electron carrying the leftover momentum $\kappa_m$ together with $m$ phonons. This is the ``phonon-cloud'' basis and bears resemblance to the spirit in which the phonon clouds in MA approximations and its numerical implementation, GGCE are implemented~\cite{Berciu2006,Sous2021}. These kets are symmetric under permutations of $q_1\dots q_m$ because the bosonic operators commute. We expand
\begin{equation}
H^{n}\ket{k}
=\sum_{m=0}^{n}\frac{1}{m!\,N^{m/2}}
\sum_{q_1\cdots q_m}
f^{(n)}_{m}(k;q_1,\dots,q_m)\,
\ket{k;q_1,\dots,q_m},
\label{eq:expansion}
\end{equation}
with the amplitudes $f^{(n)}_{m}$ chosen \emph{totally symmetric} in $q_1\dots q_m$. Without the symmetry requirement the coefficients are not unique: since $\ket{k;q_1,q_2}=\ket{k;q_2,q_1}$, only the sum $A(q_1,q_2)+A(q_2,q_1)$ of two coefficients is physically meaningful, and any antisymmetric part could be added at will. The factor $1/m!$ compensates the $m!$-fold over-counting of each unordered phonon configuration by the ordered sum $\sum_{q_1\cdots q_m}$.  An equivalent, convention-free characterization is
\begin{equation}
f^{(n)}_{m}(k;q_1,\dots,q_m)=N^{m/2}\,
\bra{k;q_1,\dots,q_m}H^{n}\ket{k},
\label{eq:f_meaning}
\end{equation}
which holds for arbitrary momenta, coincident ones included. Because $\braket{k}{k;q_1\dots q_m}=\delta_{m,0}$, the projection in Eq.~\eqref{eq:M_T0} collapses Eq.~\eqref{eq:expansion} to a single term,
\begin{equation}
M_n(k)=f^{(n)}_{0}(k),
\label{eq:Mn_is_f0}
\end{equation}
with base cases
\begin{equation}
f^{(0)}_{0}=1,\qquad f^{(0)}_{m>0}=0,
\qquad f^{(n)}_{m}\equiv 0\ \ \text{for}\ m>n .
\label{eq:base}
\end{equation}
The lower-order $f^{(j)}_{0}$ with $j<n$ do appear, but only inside the recursive evaluation of $f^{(n)}_{0}$.

\subsection{Recurrence relations at $T=0$}
\label{sec:recur_T0}

Acting with Eq.~\eqref{eq:H} on a basis state, and using
$b_p\,b^{\dagger}_{q_1}\cdots b^{\dagger}_{q_m}\ket{0}
=\sum_i\delta_{p,q_i}\,b^{\dagger}_{q_1}\cdots\widehat{b^{\dagger}_{q_i}}\cdots b^{\dagger}_{q_m}\ket{0}$ (a hat marks an omitted operator; the identity holds for coincident momenta as well), one finds
\begin{align}
H\ket{k;q_1\dots q_m}
={}&\Bigl[\varepsilon_{\kappa_m}+\sum_{i=1}^{m}\Omega(q_i)\Bigr]
\ket{k;q_1\dots q_m}
\nonumber\\
&+\frac{1}{\sqrt N}\sum_{p}g(\kappa_m,p)\,\ket{k;q_1\dots q_m,p}
\nonumber\\
&+\frac{1}{\sqrt N}\sum_{i=1}^{m}g(\kappa_m,-q_i)\,
\ket{k;q_1\dots\widehat{q_i}\dots q_m}.
\label{eq:Haction}
\end{align}
The three lines preserve, raise, and lower the phonon number by one.  Inserting Eq.~\eqref{eq:expansion} into $H^{n}\ket{k}=H\bigl(H^{n-1}\ket{k}\bigr)$ and collecting the coefficient of a fixed $\ket{k;q_1\dots q_m}$, three sources contribute: the diagonal action of $H$ on the same $m$-phonon component of $H^{n-1}\ket{k}$; emission from the $(m-1)$-phonon component, where any one of the $m$ present phonons may be the one just created; and absorption from the $(m+1)$-phonon component, integrated over the momentum of the phonon just removed. Consequently, this gives the central result of this section,
\begin{align}
f^{(n)}_{m}(k;q_1\dots q_m)
={}&\Bigl[\varepsilon_{\kappa_m}+\sum_{i=1}^{m}\Omega(q_i)\Bigr]
f^{(n-1)}_{m}(k;q_1\dots q_m)
\nonumber\\[2pt]
&+\sum_{i=1}^{m} g\bigl(\kappa_m+q_i,\;q_i\bigr)\,
f^{(n-1)}_{m-1}\bigl(k;\{q\}\!\setminus\! q_i\bigr)
\nonumber\\[2pt]
&+\Bigl\langle g\bigl(\kappa_m-p,\;-p\bigr)\,
f^{(n-1)}_{m+1}\bigl(k;q_1\dots q_m,p\bigr)\Bigr\rangle_{p} ,
\label{eq:recurrence_T0}
\end{align}
to be iterated from the base cases~\eqref{eq:base} down to $M_n=f^{(n)}_0$. Every vertex argument in Eq.~\eqref{eq:recurrence_T0} is fixed by momentum conservation across the last application of $H$: in line two that step \emph{creates} $q_i$, so the electron was at $\kappa_{m-1}=\kappa_m+q_i$ beforehand; in line three it \emph{annihilates} $p$, so the electron was at $\kappa_{m+1}=\kappa_m-p$. For the Holstein model both vertices reduce to the constant $g$; for Peierls, bond-Peierls, or breathing-mode couplings the two arguments differ and must be kept. Every term on the right of Eq.~\eqref{eq:recurrence_T0} carries the superscript $n-1$; there is no coupling between different $m$ at the same $n$. Together with Eq.~\eqref{eq:base} this guarantees termination after exactly $n$ levels.  Descending $j$ levels from $(n,m{=}0)$ one reaches states with $m\le j$; combined with $f^{(n-j)}_{m}=0$ for $m>n-j$ this gives
\begin{equation}
m\le\min(j,\,n-j)\le\Bigl\lfloor \tfrac{n}{2}\Bigr\rfloor ,
\label{eq:cone}
\end{equation}
so $M_n$ receives contributions from at most $\lfloor n/2\rfloor$ phonon lines. Since each line consumes two vertices, $M_n$ is a polynomial in the coupling containing only even powers, of degree $2\lfloor n/2\rfloor$. This is the diagram-order bound mentioned originally in Ref.\cite{Goodvin2006}, obtained organically. The recurrence generates exactly the finite set of diagrams that reach $M_n$, re-summed into closed form.

Dispersive phonons change the diagonal energy $m\Omega\to\sum_i\Omega(q_i)$; the phonon momenta were already being accounted in $\varepsilon_{\kappa_m}$. Its physical consequence is that electron and phonon momenta become correlated inside the averages, producing cross terms such as $\langle\varepsilon_{k-q}\Omega(q)\rangle_q$ from $M_4$ onward.

\subsection{The finite temperature recurrence relations}
\label{sec:TFD}

At $T>0$ the initial phonon state configurations are thermally excited, so a vertex can remove a phonon that was never created by an earlier vertex. Equation~\eqref{eq:M_shifted} shows the added complication, every intermediate state is weighted by its \emph{excitation} energy relative to the initial configuration. Rather than keeping the thermal trace and expanding, we use the thermofield-dynamics (TFD) construction of Takahashi and Umezawa~\cite{TakahashiUmezawa1975,Umezawa1982}, which maps a thermal trace onto a pure-state expectation value in a doubled Hilbert space. The transformation has recently been used to extend zero-temperature polaron solvers to finite temperature, both in multiconfigurational (Davydov-ansatz) dynamics of vibronic and polaron models~\cite{BorrelliGelin2016,ChenZhao2017} and in the generalized Green's function cluster expansion (GGCE) of Sous and coworkers~\cite{Sous2021,Carbone2023ggce,Carbone2026tfd} (for details, refer to Appendices~2.D--2.E of Carbone's thesis~\cite{CarboneThesis2021}).

Within TFD, in the doubled space the moments are again \emph{vacuum} expectation values of powers of a new effective Hamiltonian, so the construction of Secs.~\ref{sec:basis} and \ref{sec:recur_T0} applies verbatim with two boson species. As it is to be expected, handling two boson species increases the complexity of the problem \cite{MonaCovaci2007}, however the recurrence remains exact and efficient.

We add one tilde boson $\tb_q$ for each phonon mode, with $[\tb_q,\tb^{\dagger}_{q'}]=\delta_{qq'}$ and all other commutators zero, and define $\tilde H_{\rm ph}=\sum_q\Omega(q)\tb^{\dagger}_q\tb_q$. The thermal vacuum is
\begin{equation*}
\ket{0(\beta)}=\sum_\alpha\sqrt{p_\alpha}\,\ket{\alpha}\otimes\ket{\tilde\alpha}
=e^{-iG}\ket{0,\tilde 0},
\end{equation*}
\begin{equation}
    G=-i\sum_q\theta_q\bigl(b_q\tb_q-b^{\dagger}_q\tb^{\dagger}_q\bigr),
\label{eq:thermal_vacuum}
\end{equation}
with
\begin{equation}
\tanh\theta_q=e^{-\beta\Omega(q)/2},\quad
\cosh^{2}\theta_q=1+n_q,\quad
\sinh^{2}\theta_q=n_q,
\label{eq:theta}
\end{equation}
and $n_q=[e^{\beta\Omega(q)}-1]^{-1}$ the Bose occupation. Here $\ket{\tilde\alpha}$ has the same occupation numbers as $\ket{\alpha}$ and $\ket{0,\tilde0}$ is the vacuum of all $b_q$, $\tb_q$, and of the electron; $G$ is Hermitian. For every operator $O$ of the physical space, $\bra{0(\beta)}O\ket{0(\beta)}=\mathrm{Tr}[\rho\,O]$.

\paragraph*{Moments as vacuum expectation values.} The key step is to identify the operator whose powers give the moments. Let $\Hhat\equiv H-\tilde H_{\rm ph}$. Since $\tilde H_{\rm ph}$ commutes with $H$ and with $c^{\dagger}_k$, and $\tilde H_{\rm ph}\ket{\tilde\alpha}=E_\alpha\ket{\tilde\alpha}$,
\begin{equation}
\Hhat^{\,n}c^{\dagger}_k\ket{0(\beta)}
=\sum_\alpha\sqrt{p_\alpha}\,\Bigl[(H-E_\alpha)^{n}c^{\dagger}_k\ket{\alpha}\Bigr]\otimes\ket{\tilde\alpha},
\end{equation}
and orthogonality of the tilde states, $\braket{\tilde\alpha'}{\tilde\alpha}=\delta_{\alpha\alpha'}$, turns Eq.~\eqref{eq:M_shifted} into
\begin{equation}
M_n(k,T)=\bra{0(\beta)}\,c_k\,\Hhat^{\,n}\,c^{\dagger}_k\,\ket{0(\beta)}.
\label{eq:M_TFD}
\end{equation}
 Evidently, $\Hhat\ket{0(\beta)}=0$. This is the finite-$T$ analogue of $H\ket{0}=0$; $\Hhat=H-\tilde H$ is the standard generator of time evolution in TFD~\cite{Umezawa1982}. Equation~\eqref{eq:M_TFD} is a pure-state expectation value. 

Because $e^{iG}$ commutes with $c_k$, Eq.~\eqref{eq:M_TFD} becomes
\begin{equation}
M_n(k,T)=\bra{k;0,\tilde0}\,\Htheta^{\,n}\,\ket{k;0,\tilde0},
\qquad
\Htheta\equiv e^{iG}\Hhat e^{-iG},
\label{eq:M_theta}
\end{equation}
with $\ket{k;0,\tilde0}\equiv c^{\dagger}_k\ket{0,\tilde0}$. Using $e^{iG}b_qe^{-iG}=\cosh\theta_q\,b_q+\sinh\theta_q\,\tb^{\dagger}_q$ and $e^{iG}\tb_qe^{-iG}=\cosh\theta_q\,\tb_q+\sinh\theta_q\,b^{\dagger}_q$, the invariance of $b^{\dagger}_qb_q-\tb^{\dagger}_q\tb_q$, and $\theta_q=\theta_{-q}$, one obtains the general TFD Hamiltonian
\begin{align}
\Htheta={}&\sum_k\varepsilon_k\,c^{\dagger}_kc_k
+\sum_q\Omega(q)\bigl(b^{\dagger}_qb_q-\tb^{\dagger}_q\tb_q\bigr)
\nonumber\\
&+\frac{1}{\sqrt N}\sum_{k,q}g(k,q)\,c^{\dagger}_{k-q}c_k
\Bigl[\cosh\theta_q\bigl(b^{\dagger}_q+b_{-q}\bigr)
\nonumber\\
&\qquad\qquad\qquad\quad
+\sinh\theta_q\bigl(\tb^{\dagger}_{-q}+\tb_q\bigr)\Bigr].
\label{eq:Htheta}
\end{align}
Hermiticity of $H$ requires Eq.~\eqref{eq:hermiticity} and is inherited unchanged. For $\theta_q\to0$ the tilde sector decouples and Eq.~\eqref{eq:Htheta} reduces to Eq.~\eqref{eq:H}.

The operators in Eq.~\eqref{eq:Htheta} create quasiparticles relative to the thermal state, and their interpretation is physical. $b^{\dagger}_q$ adds a phonon to the bath with amplitude $\sqrt{1+n_q}$. $\tb^{\dagger}_q$ \emph{removes} a phonon from the bath with amplitude $\sqrt{n_q}$: in the original frame the state $\tb^{\dagger}_p\ket{0,\tilde0}$ is $b_p\ket{0(\beta)}/\sinh\theta_p$. A tilde phonon is a hole in the bath; it has energy $-\Omega(q)$ and momentum $-q$. The latter is consistent with the conserved total momentum of the doubled space,
\begin{equation}
\hat P=\sum_k k\,c^{\dagger}_kc_k+\sum_q q\bigl(b^{\dagger}_qb_q-\tb^{\dagger}_q\tb_q\bigr)=P-\tilde P ,
\label{eq:Phat}
\end{equation}
which satisfies $[\hat P,\Htheta]=0$ (modulo a reciprocal-lattice vector) and $[\hat P,G]=0$. Equation~\eqref{eq:Phat} is the usual TFD rule $\hat O=O-\tilde O$ for the generators of the doubled space~\cite{Umezawa1982,Balachandran2010}; 

Since $\hat P\ket{k;0,\tilde0}=k\ket{k;0,\tilde0}$, the state $\Htheta^{\,n}\ket{k;0,\tilde0}$ stays in the sector $\hat P=k$. The basis~\eqref{eq:basis} becomes
\begin{align}
\ket{k;\{q\};\{\tq\}}&\equiv c^{\dagger}_{\kappa}\,b^{\dagger}_{q_1}\cdots b^{\dagger}_{q_m}\,\tb^{\dagger}_{\tq_1}\cdots\tb^{\dagger}_{\tq_{\tm}}\ket{0,\tilde0},
\nonumber\\
\kappa&\equiv k-\sum_{i=1}^{m}q_i+\sum_{j=1}^{\tm}\tq_j ,
\label{eq:basis_T}
\end{align}
with $\{q\}=q_1\dots q_m$ physical and $\{\tq\}=\tq_1\dots\tq_{\tm}$ tilde quanta; the electron loses momentum $q$ when a physical phonon is emitted and gains $\tq$ when a bath phonon is absorbed. The expansion~\eqref{eq:expansion} and the overlap definition~\eqref{eq:f_meaning} generalize to
\begin{align}
\Htheta^{\,n}\ket{k}
&=\sum_{m+\tm\le n}
\frac{1}{m!\,\tm!\,N^{(m+\tm)/2}}
\nonumber\\
&\quad\times\sum_{\{q\},\{\tq\}}
f^{(n)}_{m,\tm}\bigl(k;\{q\};\{\tq\}\bigr) \ket{k;\{q\};\{\tq\}},
\nonumber\\
f^{(n)}_{m,\tm}
&=N^{(m+\tm)/2}
\bra{k;\{q\};\{\tq\}}\Htheta^{\,n}\ket{k}.
\label{eq:expansion_T}
\end{align}
where $f^{(n)}_{m,\tm}$ is symmetric in $\{q\}$ and in $\{\tq\}$ separately. Because $b$ and $\tb$ commute, the multiset argument of Sec.~\ref{sec:basis} applies to each species, and Eq.~\eqref{eq:expansion_T} holds for arbitrary momenta.

The action of $\Htheta$ on a basis state now has five pieces: the diagonal one, with energy $\varepsilon_\kappa+\sum_i\Omega(q_i)-\sum_j\Omega(\tq_j)$; creation and annihilation of a physical phonon, each weighted by $\cosh\theta$; and creation and annihilation of a tilde quantum, each weighted by $\sinh\theta$. Mirroring  Sec.~\ref{sec:recur_T0}, and using Eq.~\eqref{eq:hermiticity} to turn the adjoint action into vertex arguments, we get the finite-temperature recurrence which is the central result of this section:
\begin{widetext}
\begin{align}
f^{(n)}_{m,\tm}\bigl(k;\{q\};\{\tq\}\bigr)={}&
\Bigl[\varepsilon_{\kappa}+\sum_{i=1}^{m}\Omega(q_i)-\sum_{j=1}^{\tm}\Omega(\tq_j)\Bigr]\,
f^{(n-1)}_{m,\tm}\bigl(k;\{q\};\{\tq\}\bigr)
\nonumber\\
&+\sum_{i=1}^{m}\cosh\theta_{q_i}\;g\bigl(\kappa+q_i,\;q_i\bigr)\;
f^{(n-1)}_{m-1,\tm}\bigl(k;\{q\}\!\setminus\! q_i;\{\tq\}\bigr)
+\sum_{j=1}^{\tm}\sinh\theta_{\tq_j}\;g\bigl(\kappa-\tq_j,\;-\tq_j\bigr)\;
f^{(n-1)}_{m,\tm-1}\bigl(k;\{q\};\{\tq\}\!\setminus\!\tq_j\bigr)
\nonumber\\
&+\Bigl\langle\cosh\theta_{p}\;g\bigl(\kappa-p,\;-p\bigr)\;
f^{(n-1)}_{m+1,\tm}\bigl(k;\{q\},p;\{\tq\}\bigr)\Bigr\rangle_{p}
+\Bigl\langle\sinh\theta_{p}\;g\bigl(\kappa+p,\;p\bigr)\;
f^{(n-1)}_{m,\tm+1}\bigl(k;\{q\};\{\tq\},p\bigr)\Bigr\rangle_{p} ,
\label{eq:recurrence_T}
\end{align}
\end{widetext}
with
\begin{equation}
f^{(0)}_{0,0}=1,\quad f^{(n)}_{m,\tm}\equiv0\ \ \text{for}\ m+\tm>n,\quad
M_n(k,T)=f^{(n)}_{0,0}(k).
\label{eq:base_T}
\end{equation}
It is trivial to see that for $\theta_q\to0$ Eq.~\eqref{eq:recurrence_T} collapses term by term onto Eq.~\eqref{eq:recurrence_T0}.

We can choose to remove the hyperbolic functions by defining the rescaled amplitudes:
\begin{equation}
P^{(n)}_{m,\tm}\equiv\Bigl[\prod_{i=1}^{m}\cosh\theta_{q_i}\prod_{j=1}^{\tm}\sinh\theta_{\tq_j}\Bigr]\,f^{(n)}_{m,\tm}.
\label{eq:P_def}
\end{equation}
Multiplying Eq.~\eqref{eq:recurrence_T} by the bracket, the discrete sums acquire the weights $\cosh^{2}\theta_{q_i}=1+n_{q_i}$ and $\sinh^{2}\theta_{\tq_j}=n_{\tq_j}$, while the two zone averages carry a bare vertex:
\begin{equation}
\begin{aligned}
P^{(n)}_{m,\tm}
&=\Bigl[\varepsilon_{\kappa}
+\sum_i\Omega(q_i)
-\sum_j\Omega(\tq_j)\Bigr]P^{(n-1)}_{m,\tm}
\\
&\quad+\sum_{i=1}^{m}(1+n_{q_i})\,
g\bigl(\kappa+q_i,q_i\bigr)
\\
&\qquad\times P^{(n-1)}_{m-1,\tm}
\bigl(\{q\}\!\setminus\!q_i;\{\tq\}\bigr)
\\
&\quad+\sum_{j=1}^{\tm}n_{\tq_j}\,
g\bigl(\kappa-\tq_j,-\tq_j\bigr)
\\
&\qquad\times P^{(n-1)}_{m,\tm-1}
\bigl(\{q\};\{\tq\}\!\setminus\!\tq_j\bigr)
\\
&\quad+\Bigl\langle g\bigl(\kappa-p,-p\bigr)\,
P^{(n-1)}_{m+1,\tm}
\bigl(\{q\},p;\{\tq\}\bigr)\Bigr\rangle_p
\\
&\quad+\Bigl\langle g\bigl(\kappa+p,p\bigr)\,
P^{(n-1)}_{m,\tm+1}
\bigl(\{q\};\{\tq\},p\bigr)\Bigr\rangle_p .
\end{aligned}
\label{eq:recurrence_P}
\end{equation}

with the base cases of Eq.~\eqref{eq:base_T} and $M_n=P^{(n)}_{0,0}$. 

The lowest orders read, for any vertex in class~\eqref{eq:H}:
\begin{align}
M_2(k,T)&=\varepsilon_k^{2}+\bigl\langle(1+2n_q)\,|g(k,q)|^{2}\bigr\rangle_q ,
\label{eq:M2}\\
M_3(k,T)&=\varepsilon_k^{3}+\Bigl\langle|g(k,q)|^{2}\bigl[(1+2n_q)(2\varepsilon_k+\varepsilon_{k-q})+\Omega(q)\bigr]\Bigr\rangle_q .
\label{eq:M3}
\end{align}
The term $\Omega(q)=(1+n_q)\Omega(q)+n_q[-\Omega(q)]$ in $M_3$ carries no thermal factor. For Einstein phonons, Eqs.~\eqref{eq:M2}--\eqref{eq:M3} are the general-vertex results of Ref.~\cite{Mitric2026}, and for the Holstein model they reduce to the expressions in Refs.~\cite{Kornilovitch2002,Bonca2019}.

\subsection{Accessible polaron models}
\label{sec:reach}

Table~\ref{tab:vertices} lists the couplings that can be accessed through the existing generic form of the vertex, in the convention of Eq.~\eqref{eq:H}, together with the second moment that Eq.~\eqref{eq:M2} assigns to each. Here, we first note that the the factor $-2i\sin(q/2)$ that converts a site coordinate into a bond-length change is what separates the site-Peierls from the bond-Peierls coupling and the breathing-mode from the Holstein coupling; it removes the $q=0$ coupling and changes the $k$ dependence of $M_2$.  Second, the Edwards fermion--boson coupling~\cite{Edwards2006,Alvermann2007} is \emph{not} of the form~\eqref{eq:H}: a hop from $i$ to $j$ creates a boson at $i$ or removes one at $j$, so the emission and absorption vertices differ, $g_e(k,q)=-2t_b\cos(k-q)$ and $g_a(k,q)=-2t_b\cos k$. The recurrence accommodates this with no change of structure---the creation lines carry $g_e$ and the annihilation lines $g_a$. 

\begin{table*}[t]
\caption{\label{tab:vertices}Linear electron--phonon couplings of single-orbital lattice models in the convention of Eq.~\eqref{eq:H}, in $d=1$ with $\varepsilon_k=-2t\cos k$. $X_r=N^{-1/2}\sum_qe^{-iqr}(b^{\dagger}_q+b_{-q})$ is the phonon coordinate at a site (integer $r$) or bond (half-integer $r$); the real-space column omits $\sum_i$. The last column is $M_2-\varepsilon_k^{2}$ for Einstein phonons, with $\nB$ the Bose occupation. The Edwards coupling is listed with its separate emission and absorption vertices.}
\small
\begin{ruledtabular}
\begin{tabular}{llllc}
Coupling & Real space & $g(k,q)$ & $\lambda$ & $M_2-\varepsilon_k^{2}$ \\
\hline
Holstein~\cite{Holstein1959} & $g\,n_iX_i$ & $g$ & $g^{2}/(2t\Omega)$ & $g^{2}(2\nB+1)$ \\
Site-Peierls (SSH)~\cite{Marchand2010} & $g\,(c^{\dagger}_ic_{i+1}+\mathrm{h.c.})(X_{i+1}-X_i)$ & $-4ig\sin\frac q2\cos(k-\frac q2)$ & $2g^{2}/(t\Omega)$ & $2g^{2}(1+2\sin^{2}k)(2\nB+1)$ \\
Bond-Peierls~\cite{Carbone2021bond} & $g\,(c^{\dagger}_ic_{i+1}+\mathrm{h.c.})X_{i+1/2}$ & $2g\cos(k-\frac q2)$ & $g^{2}/(t\Omega)$ & $2g^{2}(2\nB+1)$ \\
Breathing mode~\cite{Lau2007} & $g\,n_i(X_{i+1/2}-X_{i-1/2})$ & $-2ig\sin\frac q2$ & $g^{2}/(t\Omega)$ & $2g^{2}(2\nB+1)$ \\
Edwards~\cite{Alvermann2007} 
& $-t_b\,c^{\dagger}_jc_i(b^{\dagger}_i+b_j)$ 
& $\begin{aligned}[t]
g_e&=-2t_b\cos(k-q),\\
g_a&=-2t_b\cos k
\end{aligned}$
& --- 
& $2t_b^{2}[1+\nB(1+2\cos^{2}k)]$ \\
\end{tabular}
\end{ruledtabular}
\end{table*}

 The recurrence produces a finite answer only if the momentum averages it contains converge, and this is a property of the model, not of the method. From Eq.~\eqref{eq:recurrence_P}, we can see that each closed phonon line carries two vertices at the same momentum, one factor $2n_q+1$ or $1$, and a bounded polynomial in $\varepsilon$ and $\Omega$, so the convergence of every moment is controlled by that of
\begin{equation}
I(k,T)=\int_{\rm BZ}\frac{d^{d}q}{(2\pi)^{d}}\,|g(k,q)|^{2}\coth\frac{\beta\Omega(q)}{2},
\label{eq:I_criterion}
\end{equation}
which is $M_2-\varepsilon_k^{2}$ itself: if $I$ diverges, $M_2$ diverges, and if $I$ converges with $|g|^{2}$ replaced by $\max_k|g|^{2}$, all moments are finite. Near a point $q_*$ where $\Omega$ vanishes or $g$ is singular, write $|g|^{2}\simeq C|p|^{a}$ and $\Omega\simeq c|p|^{b}$ with $p=q-q_*$; since $\coth x\simeq1/x$ for small $x$, the integral converges if
\begin{equation}
a+d>0\ \ (T=0\ \text{or}\ b=0),\qquad a-b+d>0\ \ (T>0,\ b>0),
\label{eq:power_counting}
\end{equation}
with a logarithmic divergence when the left-hand side vanishes. This lets us draw the following conclusions:

(a) On a lattice with gapped phonons---Einstein or dispersive optical with $\Omega(q)>0$ everywhere---all couplings of Table~\ref{tab:vertices} give finite moments at all temperatures. 

(b) For acoustic phonons ($b=1$) the finite-temperature moments are finite only if the coupling vanishes with $q$. For the ``acoustic Holstein''  model, it has $a=0$ and its $M_2$ diverges logarithmically at $T>0$ in $d=1$; Conversely, for the acoustic SSH coupling~\cite{MarchandThesis2011}, all their moments are finite at any temperature in any dimension. 

In practise, the difficulty of calculating high orders of moments does depend on the model chosen. Those with complex vertices (such as the SSH case) have more terms and are more expensive in terms of computer memory, compared to simpler models such as the Holstein. Similarly, Einstein phonons are much cheaper to evaluate due to the momentum averages over the BZ becoming independent of phonon momenta.

\section{Sum-rule validation of existing analytical methods}
\label{sec:validation}

The most immediate application of a set of exact moments is  testing the accuracy of analytical methods. An approximate Green's function (obtained, say through a self-energy) has its own large-$z$ expansion, and comparing its coefficients with the exact $M_n$, order by order and in closed form, which sum rules the approximation satisfies and which diagram is responsible for the first failure. Because the diagram-order bound of Sec.~\ref{sec:recur_T0} ties $M_n$ to diagrams with at most $\lfloor n/2\rfloor$ phonon lines, the first failing moment also tells one the order in $g^{2}$ at which the approximation ceases to be exact. We do so here for the one-dimensional Holstein and Peierls models at zero and finite temperature. The closed-form expansions themselves are too long to print; they are available, together with the exact moments, in the accompanying data files.

The one-shot Born approximation, with the lowest-order self-energy inserted once into Dyson's equation, and its self-consistent version (SCBA) can only be considered valid in the weak-coupling regime~\cite{Migdal1958,EngelsbergSchrieffer1963, Goodvin2006}. BA/SCBA reproduces the sum rules exactly for $n<4$; the first error of both is $O(g^{4})$ at $M_4$, from the omitted fourth-order diagrams. Our Holstein and Peierls moments confirm this at every temperature. The result follows directly from the diagram-order bound: a self-energy that is exact through order $g^{2}$ is guaranteed to be exact through $M_3$, and not beyond.

Ref. \cite{Mitric2023,MitricThesis2023}, showed that the second-order cumulant expansion (CE) fares better for the Holstein model: it reproduces the moments exactly through $M_4$, and the first discrepancy is $M_5^{\rm CE}(k)-M_5(k)=-2g^{4}\varepsilon_k\,(2\nB+1)^{2}$ . For a momentum-dependent vertex the CE loses another order: for the Peierls model its moments are exact only through $M_3$, with a generic $O(g^{4})$ error at $M_4$ that is proportional to $(2\nB+1)^{2}$ ~\cite{Mitric2026}. The higher CE moments tabulated in the Supplemental Material of Ref.~\cite{Mitric2026}; our recurrence supplies the exact Peierls moments and the two agree through $M_3$ and differ from $M_4$ on, as expected.

On the other side of the spectrum, we have the momentum-average approximation and DMFT. Berciu's original argument was that MA$^{(0)}$ not only satisfies the first six Holstein sum rules exactly, but that it satisfies \emph{all} higher ones with good accuracy, because every diagram is kept and only the free-propagator momenta inside each are averaged over the Brillouin zone\cite{Berciu2006,Goodvin2006}.  MA$^{(1)}$, which keeps the one-phonon state exactly and averages only from two phonons upward, was reported to satisfy two further sum rules~\cite{BerciuGoodvin2007}. The finite-temperature extension of MA~\cite{Shannigrahi2026} do not clarify whether these claims change at $T>0$.

Expanding the MA$^{(0)}$ and MA$^{(1)}$ self-energies in $1/z$ symbolically and comparing with the exact moments, we find the following for the 1D Holstein model, with $s\equiv2\nB+1$ ($s=1$ at $T=0$) and $\Delta M_n\equiv M_n^{\rm exact}-M_n^{\rm approx}$:
\begin{align}
\text{MA}^{(0)}:&\quad \Delta M_{0\ldots5}=0,\qquad \Delta M_6=2g^{4}t^{2}s^{2},
\label{eq:dMA0}\\
\text{MA}^{(1)}:&\quad \Delta M_{0\ldots6}=0,\qquad \Delta M_7=\varepsilon_k\,g^{4}t^{2}s^{2},
\label{eq:dMA1}
\end{align}
as exact polynomial identities valid for arbitrary $t,g,\Omega>0$ and any $\nB\ge0$. Equation~\eqref{eq:dMA0} at $T=0$ is the statement $M_6^{{\rm MA}^{(0)}}=M_6-2dt^{2}g^{4}$ of Ref.~\cite{Goodvin2006} for $d=1$. Equation~\eqref{eq:dMA1} is new and corrects the published count: MA$^{(1)}$ is exact through $M_6$, not $M_7$. 

At order $g^{4}$ the self-energy contains the non-crossed/rainbow diagram and the first crossed diagram. MA$^{(0)}$ replaces the correlated average $\langle G_0(p,z-\Omega)^{2}\rangle_p$ of the non-crossed diagram by $\langle G_0(p,z-\Omega)\rangle_p^{2}$ and gives Eq.~\eqref{eq:dMA0}. MA$^{(1)}$ corrects this term exactly but does not completely capture the crossed diagram. Moreover, the crossed diagram is the one generating k-dependence in the self-energy. For the Holstein model, the MA$^{(1)}$ self-energy is local likein  dynamical mean-field theory~\cite{Ciuchi1997,Mitric2022} and both are exact through $M_6$ with the identical error $\Delta M_7=\varepsilon_kg^{4}t^{2}s^{2}$. 

At $t=0$ both MA$^{(0)}$ and MA$^{(1)}$ (including the finite-T generalization) reproduce every moment we can check ($M_0$--$M_{28}$) at all temperatures, confirming the claim from Ref. ~\cite{Shannigrahi2026}  that the finite-$T$ MA self-energy is asymptotically exact in the atomic limit; the same is true trivially at $g=0$.

Table~\ref{tab:exactness} summarizes the analytical methods for which the moment deviation can be clearly summarized. 

\begin{table}[t]
\caption{\label{tab:exactness}Highest moment reproduced exactly by common approximations for the 1D Holstein and Peierls models with Einstein phonons, at $T=0$ and $T>0$ (same order in both cases), for the spectral function at generic $k$. ``First error'' gives the leading discrepancy $M_n^{\rm exact}-M_n^{\rm approx}$ where known in closed form; $s=2\nB+1$ for the Holstein model. The multi-site MA for Peierls are not analysed here and reserved for future work.}
\begin{ruledtabular}
\begin{tabular}{lccl}
Approximation & Holstein & Peierls & First error \\
\hline
Born (one-shot Migdal)& $M_3$ & $M_3$ & $O(g^{4})$ at $M_4$ \\
SCBA / SCMA & $M_3$ & $M_3$ & $O(g^{4})$ at $M_4$ \\
Second-order CE~\cite{Mitric2023,Mitric2026} & $M_4$ & $M_3$ & $2g^{4}\varepsilon_ks^{2}$ at $M_5$ \\
MA$^{(0)}$~\cite{Goodvin2006} & $M_5$ & --- & $2g^{4}t^{2}s^{2}$ at $M_6$ \\
MA$^{(1)}$~\cite{BerciuGoodvin2007} & $M_6$ & --- & $\varepsilon_kg^{4}t^{2}s^{2}$ at $M_7$ \\
DMFT~\cite{Ciuchi1997,Mitric2022} & $M_6$ & --- & $\varepsilon_kg^{4}t^{2}s^{2}$ at $M_7$ \\
\end{tabular}
\end{ruledtabular}
\end{table}
\section{Spectral functions from the moments}
\label{sec:reconstruction}

The moments are the coefficients of the large-$z$ expansion of the Green's function, and we first considered using that expansion directly. Writing $G(k,z)=\int d\nu\,A(k,\nu)/(z-\nu)$ with $z=\omega+i\eta$ and expanding $1/(z-\nu)=z^{-1}\sum_{m\ge0}(\nu/z)^{m}$ term by term gives
\begin{equation}
G(k,z)=\sum_{m=0}^{\infty}\frac{M_m(k)}{z^{m+1}} ,
\label{eq:series}
\end{equation}
which converges for $|z|>R$ whenever $A(k,\nu)$ vanishes outside $|\nu|<R$.  However, no electron--phonon model of the form~\eqref{eq:H} has bounded spectral support. The electron couples to states with arbitrarily many phonons and arbitrarily high energy, and although the weight at energy $\nu$ falls off faster than any power, it never vanishes. Hence, the series~\eqref{eq:series} has zero radius of convergence, it is only an asymptotic expansion \cite{Schmudgen2017}. At sufficiently large $|z|$ a few terms approximate $G$ well and adding terms improves the asymptotic accuracy, but at any fixed finite $z$ adding more and more terms eventually makes the approximation worse. None of this is a limitation of the moments themselves, which fix $G$ to arbitrary accuracy as we show below; it is a limitation of summing them as a power series in $1/z$.

At the same time, the series make the contribution of the diagrams to the moments quite transparent.  Writing $G=1/[z-\varepsilon_k-\Sigma(k,z)]$ and expanding, the coefficient of $z^{-(m+1)}$ collects products of $\varepsilon_k$ and of the coefficients $C_j$ of $\Sigma=\sum_jC_j/z^{j+1}$, so that
\begin{equation}
M_m=\varepsilon_kM_{m-1}+\sum_{j=0}^{m-2}C_j\,M_{m-2-j},
\label{eq:M_from_C}
\end{equation}
with $M_0=1$. The lowest-order self-energy of the Holstein model, $\Sigma^{(1)}=g^{2}\langle G_0(p,z-\Omega)\rangle_p=g^{2}g_0(z-\Omega)$ with $g_0(z)=[z^{2}-4t^{2}]^{-1/2}$ in 1D, expands as $g_0(z-\Omega)=z^{-1}+\Omega z^{-2}+(\Omega^{2}+2t^{2})z^{-3}+\cdots$, giving $C_0=g^{2}$ and $C_1=g^{2}\Omega$ and hence $M_2=\varepsilon_k^{2}+g^{2}$ and $M_3=\varepsilon_k^{3}+2g^{2}\varepsilon_k+g^{2}\Omega$, the $T=0$ results of Eqs.~\eqref{eq:M2}--\eqref{eq:M3}. The term $2g^{2}t^{2}$ in $M_4$ is the first place where the electron hops while the phonon is present; $\langle\varepsilon_p\rangle_p=0$ kills the would-be lower-order terms. 

To determine whether the idea of reconstructing the spectral function from a finite set of moments is even a well-defined problem for the polaron, we rely on the validity of Carleman's criterion \cite{Carleman1922},
\begin{equation}
\sum_{n\ge1}M_{2n}^{-1/(2n)}=\infty .
\label{eq:carleman}
\end{equation}
For the models considered here it is satisfied. At $T=0$, $M_{2n}=\|H^{n}\ket{k}\|^{2}$, and on vectors with at most $m$ phonons the bounded dispersion, bounded phonon frequencies, and linear coupling give $\|H\phi\|\le C(m+1)\|\phi\|$ with a constant independent of $m$ and of the lattice size; since each application of $H$ raises the phonon number by at most one, $\|H^{n}\ket{k}\|\le C^{n}n!$ and $M_{2n}^{-1/(2n)}\ge1/(Cn)$, whose sum diverges. At finite temperature the same bound applies in the doubled space of Sec.~\ref{sec:TFD} too. For every lattice model with a gapped phonon branch the complete moment sequence therefore determines $A(k,\omega)$ uniquely at all temperatures.

\subsection{The continued fraction approach to the problem}
\label{sec:CF}

The classical approach is through orthogonal polynomials~\cite{Gautschi2004,ViswanathMuller1994}. Use $A(k,\cdot)$, which is a probability density in $\omega$ at every temperature, as a weight and define $\langle f,g\rangle=\int d\omega\,f(\omega)g(\omega)A(k,\omega)$. Gram--Schmidt orthogonalization of $1,\omega,\omega^{2},\dots$ produces monic polynomials $\pi_n(\omega)=\omega^{n}+\cdots$ that satisfy the three-term recurrence
\begin{equation}
\pi_{n+1}(\omega)=(\omega-a_n)\pi_n(\omega)-b_n\pi_{n-1}(\omega),\qquad \pi_0=1,\ \pi_{-1}=0,
\label{eq:three_term}
\end{equation}
with $a_n=\langle\omega\pi_n,\pi_n\rangle/\langle\pi_n,\pi_n\rangle$ and $b_n=\langle\pi_n,\pi_n\rangle/\langle\pi_{n-1},\pi_{n-1}\rangle$. Only three terms occur because $\omega\pi_m$ has degree $m+1$ and is orthogonal to $\pi_n$ for $m\le n-2$. Positivity of $A$ makes every $b_n>0$. Since $\langle\omega^{i},\omega^{j}\rangle=M_{i+j}$, every inner product is a finite combination of moments, and the counting rule follows: \emph{$M_0,\dots,M_{2N}$ fix exactly $a_0,\dots,a_{N-1}$ and $b_1,\dots,b_N$}. $N$ denotes the number of levels ; $M_0$--$M_{20}$ gives $N=10$ and $M_0$--$M_{28}$ gives $N=14$. For the Holstein model at any temperature,
\begin{align}
a_0&=\varepsilon_k,\qquad b_1=(2\nB+1)g^{2},\qquad a_1=\frac{\Omega}{2\nB+1},
\nonumber\\
b_2&=2t^{2}+2(2\nB+1)g^{2}+\frac{4\Omega^{2}\nB(\nB+1)}{(2\nB+1)^{2}},
\label{eq:ab_holstein}
\end{align}

In the normalized basis $p_n=\pi_n/\sqrt{b_1\cdots b_n}$, multiplication by $\omega$ is the tridiagonal Jacobi matrix $J$ with diagonal $a_n$ and off-diagonal $\sqrt{b_n}$. At $T=0$ this is the Lanczos algorithm in disguise~\cite{Lanczos1950,HaydockHeineKelly1972}: starting from $\ket{\psi_0}=c^{\dagger}_k\ket{0}$, the Lanczos vectors are $\ket{\psi_n}=p_n(H)\ket{\psi_0}$, $J$ is $H$ projected onto the Krylov space ${\rm span}\{\ket{\psi_0},H\ket{\psi_0},\dots,H^{N-1}\ket{\psi_0}\}$, and $a_n$, $\sqrt{b_n}$ are the site energies and hoppings of a semi-infinite chain. At $T>0$ the algebra is unchanged because $A$ is still a positive measure, and the operator-space version is Mori's recursion method~\cite{ViswanathMuller1994}. The Green's function is the $(0,0)$ element of the resolvent of $J$, and a Schur-complement step repeated level by level gives the Jacobi continued fraction
\begin{equation}
G(k,z)=\cfrac{1}{z-a_0-\cfrac{b_1}{z-a_1-\cfrac{b_2}{\ddots\;-\cfrac{b_{N-1}}{z-a_{N-1}-b_N\,\tau_N(z)}}}} ,
\label{eq:CF}
\end{equation}
where $\tau_N(z)=[(z-J^{(N)})^{-1}]_{00}$ is the Green's function of the deep part of the chain, levels $N,N+1,\dots$. Equation~\eqref{eq:CF} is exact for every positive spectral function. The first $N$ levels contain precisely the information in $M_0$--$M_{2N}$; the tail, or \emph{terminator}, $\tau_N$ is again the Stieltjes transform of a probability measure that the moments do not fix. Every admissible choice of $\tau_N$ gives a positive $A$ with the same $M_0$--$M_{2N}$. In particular, $\tau_N=0$ gives the $N$-point Gauss rule, $G(z)=\sum_{j=1}^{N}w_j/(z-x_j)$ with nodes $x_j$ the eigenvalues of the $N\times N$ Jacobi matrix and weights $w_j$ the squared first components of its eigenvectors; this discrete measure reproduces $M_0$--$M_{2N-1}$ exactly and is also the $[N-1/N]$ Pad\'e approximant of Eq.~\eqref{eq:series}. The coefficients $a_n,b_n$ are computed from the moments by the Chebyshev algorithm, a recursion in the mixed moments $\sigma_{k,l}=\langle\pi_k,\omega^{l}\rangle$~\cite{Gautschi2004}. 

The choice of terminator is where physics enters, and we have compared several~\cite{Haydock1980,BeerPettifor1984,Pinna2025}. The \emph{square-root terminator} assumes that the chain coefficients are constant beyond level $N$, $a_n=a_{N-1}$ and $b_n=b_N$, so that the deep chain reproduces itself, $\tau=1/(z-a_\infty-b_\infty\tau)$, giving $\tau(z)=[z-a_\infty-\sqrt{(z-a_\infty)^{2}-4b_\infty}]/2b_\infty$.  It is exact for a free tight-binding band ($a_n=0$, $b_n=t^{2}$ for $n\ge2$ in 1D). However, typically the polaron model coefficients grow with $n$ (the spectrum is unbounded above), so no constant terminator is exact. 

Consequently, we use a \emph{linear terminator}, which continues the last available Jacobi coefficients as
\begin{equation}
a_{N+j}=a_{N-1}+(j+1)\Delta a,\qquad
b_{N+j}=b_N+j\Delta b,\qquad j\geq0,
\end{equation}
with slopes inferred from the last few exact coefficients and chosen so that the continued $b_n$ remain positive. This follows the recursion-method strategy of replacing the unknown part of the continued fraction by a model with the observed coefficient trend~\cite{ViswanathMuller1994}. It leaves $M_0$--$M_{2N}$ unchanged. When $t=0$, the local phonon-number basis already makes the Hamiltonian tridiagonal, with diagonal matrix elements $n\Omega$ and neighbouring matrix elements $g\sqrt{n+1}$; hence $a_n=n\Omega$ and $b_n=ng^2$. Continuing these coefficients therefore reproduces the entire atomic Green's function so the spectral function with the linear tail  becomes exact at $T=0,t=0$.

Alternatively,  at finite temperature, phonon absorption also generates spectral weight below the zero-temperature ground-state energy.  We therefore use a \emph{constant-$a$/linear-$b$} continuation, $a_{N+j}=a_{N-1}$ and $b_{N+j}=b_N+j\Delta b$, with $\Delta b>0$. The construction belongs to the Gaussian-terminator family of the recursion method~\cite{ViswanathMuller1994,Giesekus1995}.

\begin{figure*}[tp]
\centering
\includegraphics[width=\textwidth]{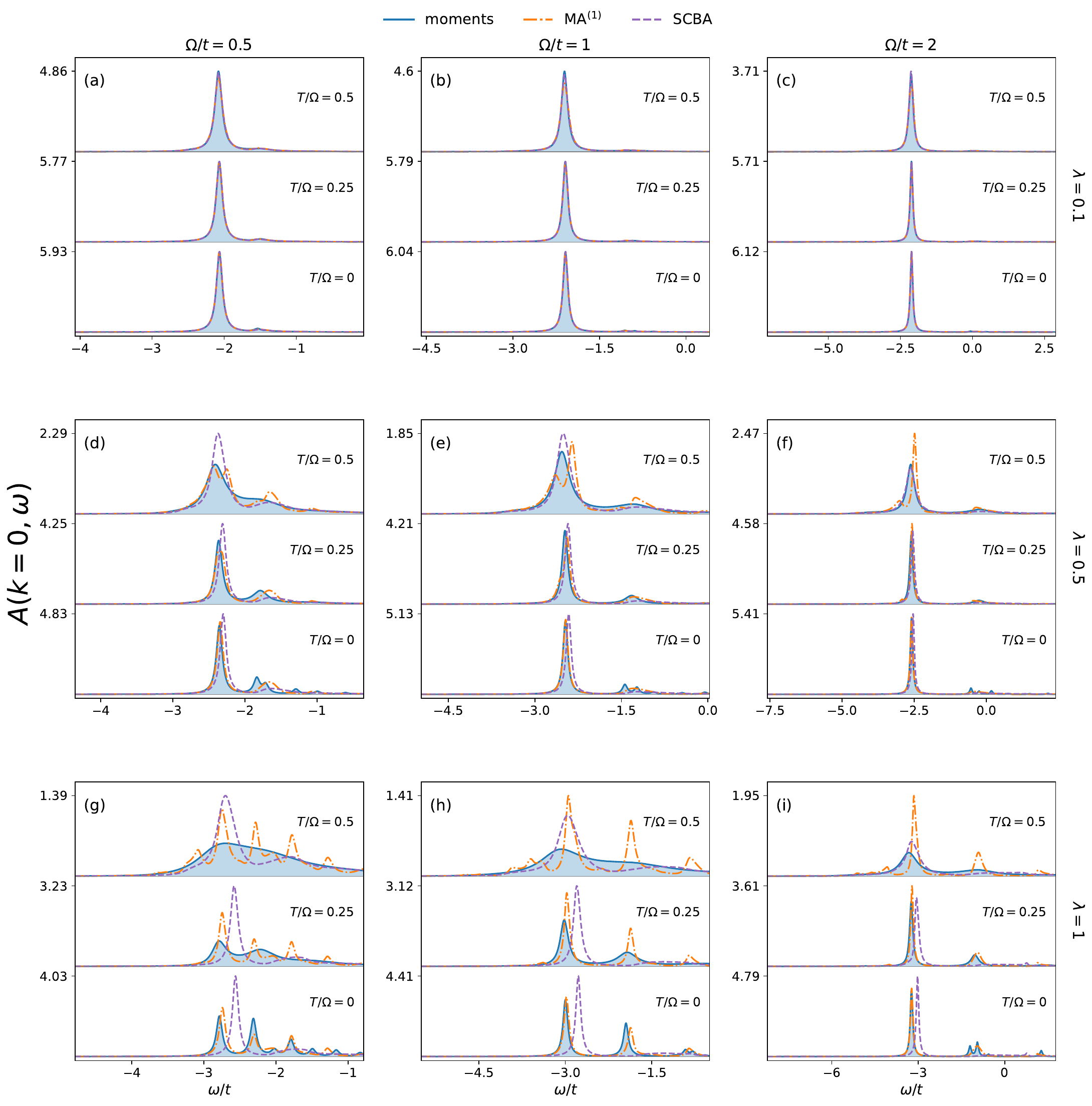}
\caption{\label{fig:spectra}Holstein spectral function at $k=0$, for $\Omega/t=0.5$, 1, 2 (columns) and $\lambda=0.1$, 0.5, 1 (rows). Each panel stacks $T/\Omega=0$, 0.25, 0.5 from bottom to top. Moments (blue solid with fill), MA$^{(1)}$ (orange dash-dotted), and SCBA (purple dashed) use the same $\eta=0.05t$.  The moment tail is linear in $a_n$ and $b_n$ at $T=0$ and constant in $a_n$, linear in $b_n$ at $T>0$.}
\end{figure*}

\begin{figure*}[tp]
\centering
\includegraphics[width=\textwidth]{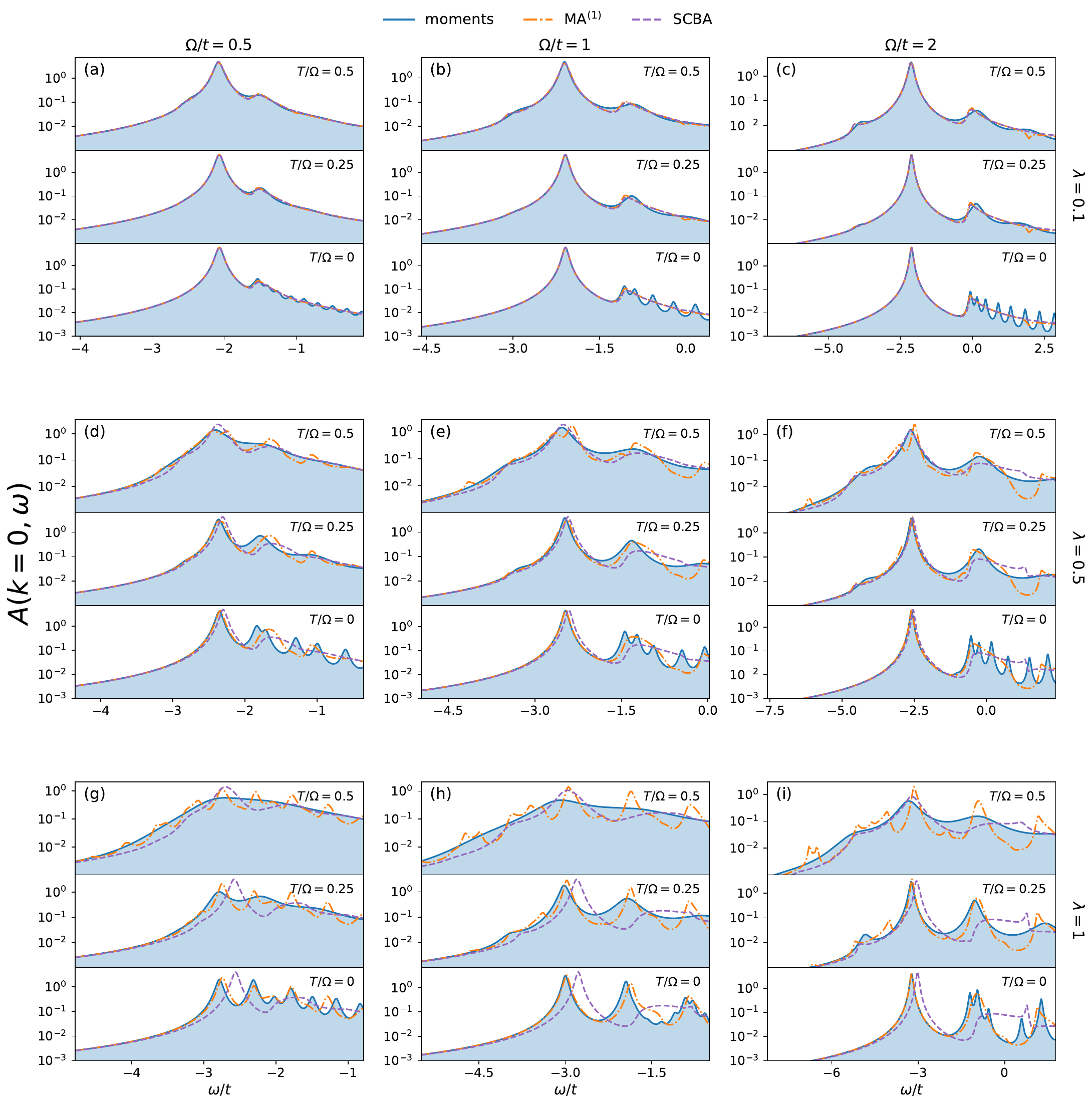}
\caption{\label{fig:hp_spectra_log}Logarithmic comparison of the same Holstein spectra as Fig.~\ref{fig:spectra}, with identical parameters. Each temperature has its own logarithmic vertical axis showing the absolute spectral function.}
\end{figure*}

\begin{figure*}[tp]
\centering
\includegraphics[width=0.94\textwidth]{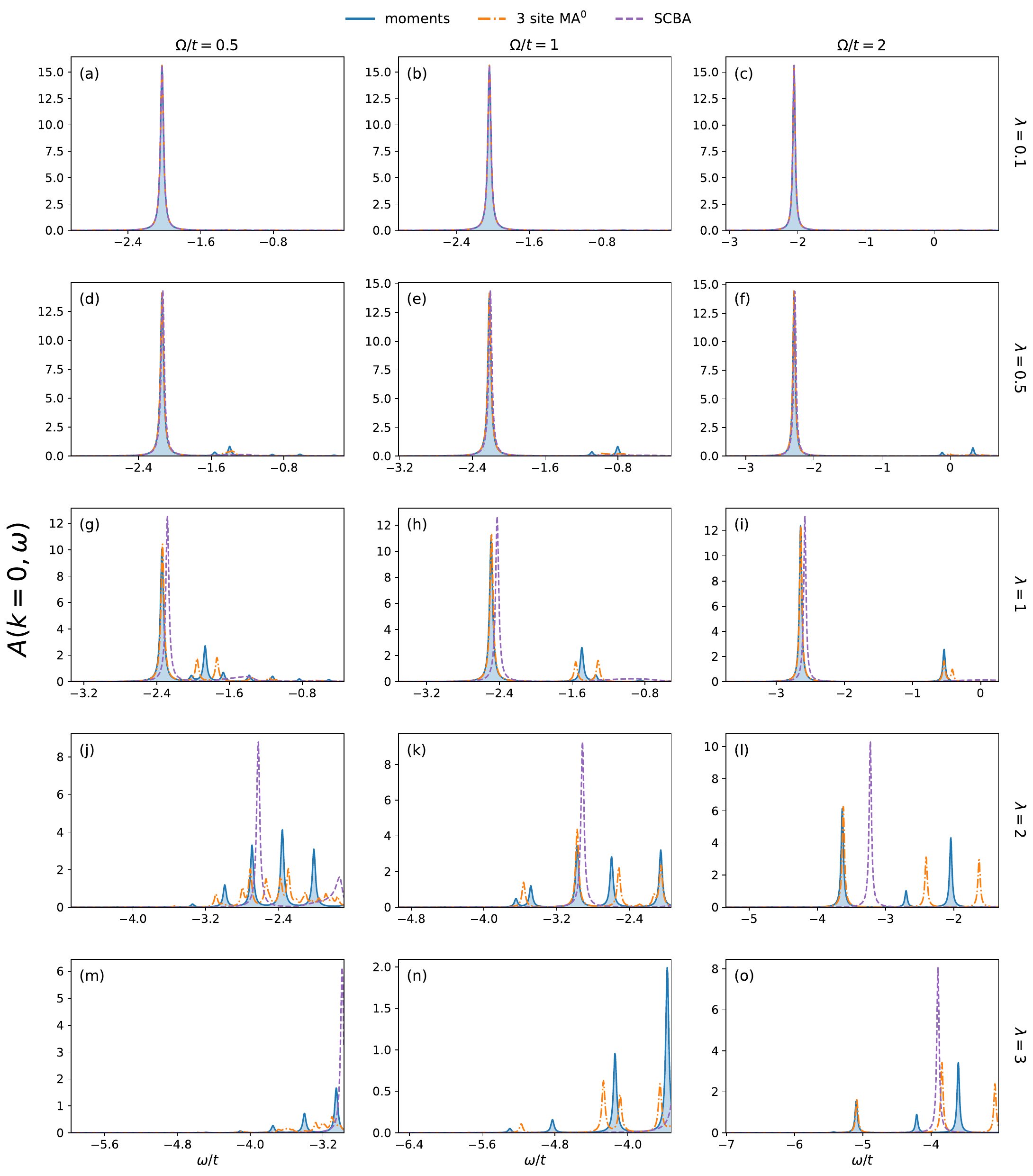}
\caption{\label{fig:ssh_spectra}SSH spectral function at $k=0$, $T=0$, for $\Omega/t=0.5$, 1, 2 (columns) and $\lambda=0.1$, 0.5, 1, 2, 3 (rows). Moments ($M_0$--$M_{23}$, blue filled), three-site MA$^{(0)}$ (orange dash-dotted), and SCBA (purple dashed) all use $\eta=0.02t$. The moment tail is linear in both $a_n$ and $b_n$. }
\end{figure*}

\begin{figure*}[tp]
\centering
\includegraphics[width=0.94\textwidth]{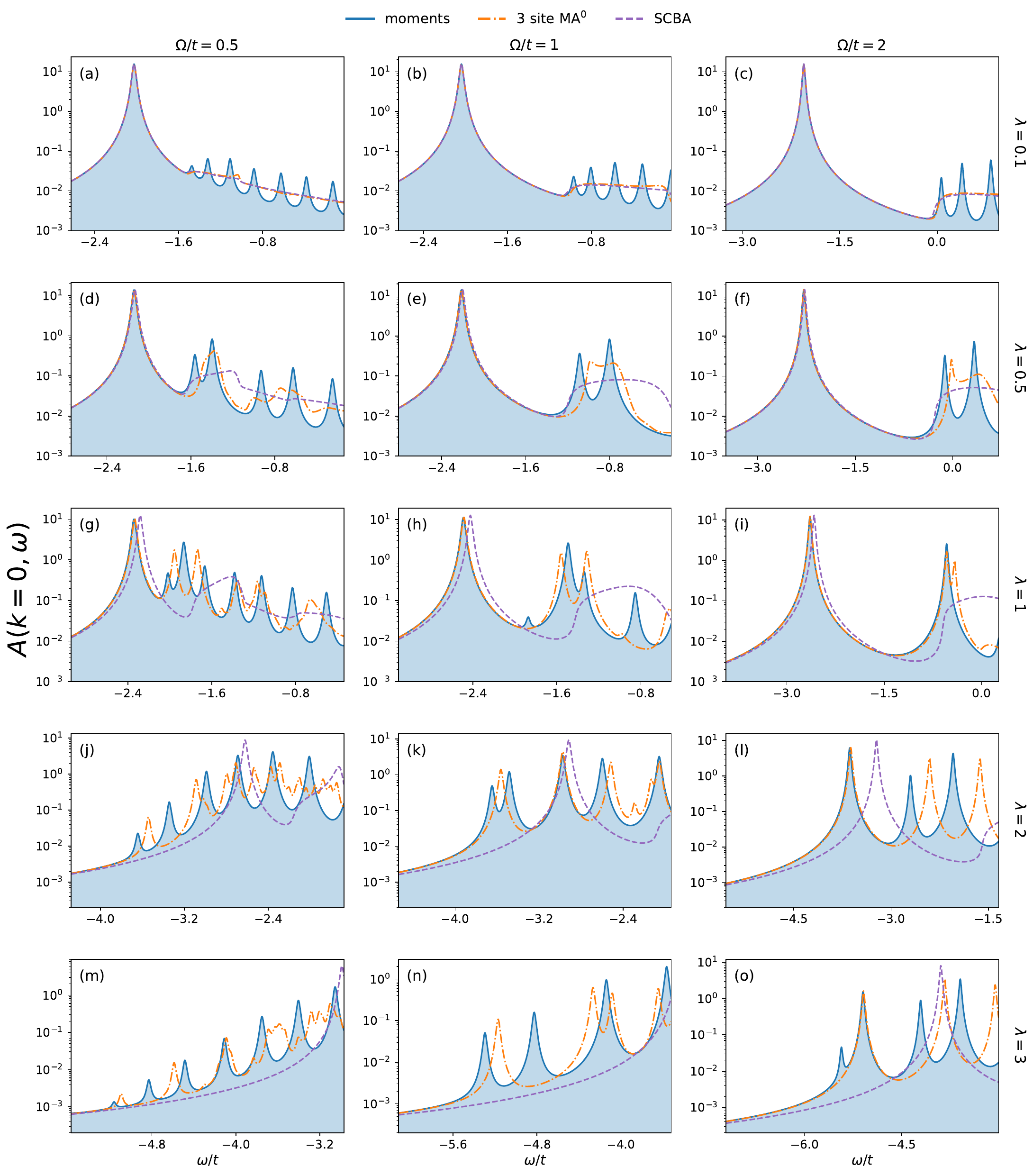}
\caption{\label{fig:ssh_spectra_log}Logarithmic view of the SSH spectra of Fig.~\ref{fig:ssh_spectra}, with the same parameter grid, curves, and $\eta=0.02t$.}
\end{figure*}
At $T\approx0$, the polaron is an isolated point of the spectrum below a gap, and the lowest Gauss node converges to it exponentially fast; the quasiparticle ground state are then essentially independent of the terminator and are fixed by the finite-set of moments to high precision. However, the incoherent continuum is not well resolved. The reason is that the high moments are dominated by the tails so in practise, even higher number of moments are needed. This ties in well with why trying to obtain the moments by integrating $\omega^{n}A(k,\omega)$ over a numerically computed spectrum is not suitable.  Spectra are rarely accurate, and a finite broadening makes matters qualitatively worse resulting in severe error propagation when constructing $a_n$,$b_n$ from the these computed moments.

\begin{figure*}[tp]
\centering
\includegraphics[width=\textwidth]{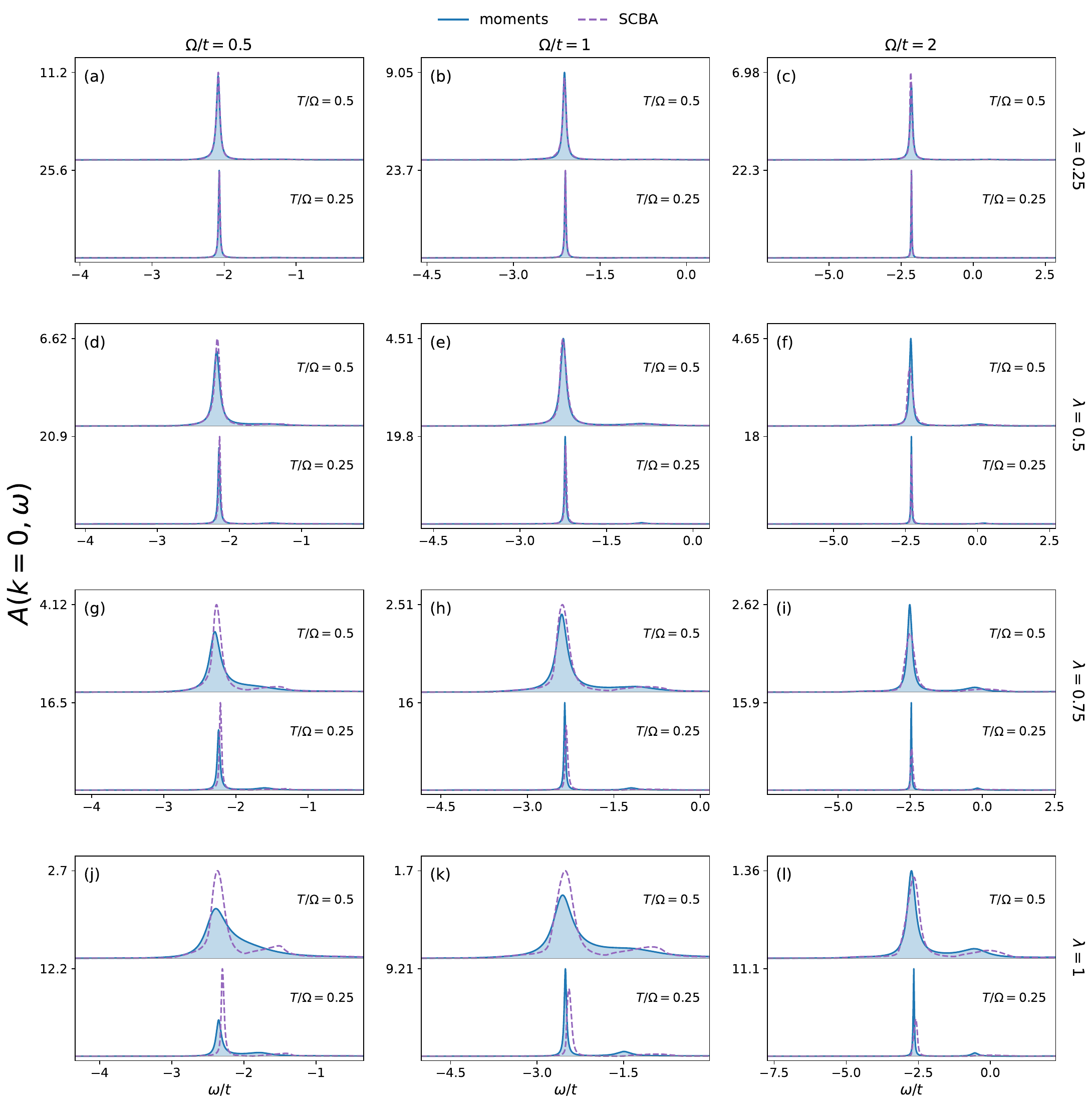}
\caption{\label{fig:ssh_thermal}Finite-temperature SSH spectra at $k=0$ for $\Omega/t=0.5$, 1, 2 (columns) and $\lambda=0.25$, 0.5, 0.75, 1 (rows). The lower and upper bands are $T/\Omega=0.25$ and 0.5. Moments ($M_0$--$M_{23}$, blue filled) use a constant-$a$/linear-$b$ tail; SCBA is purple dashed. Both use $\eta=0.01t$.}
\end{figure*}

\begin{figure*}[tp]
\centering
\includegraphics[width=\textwidth]{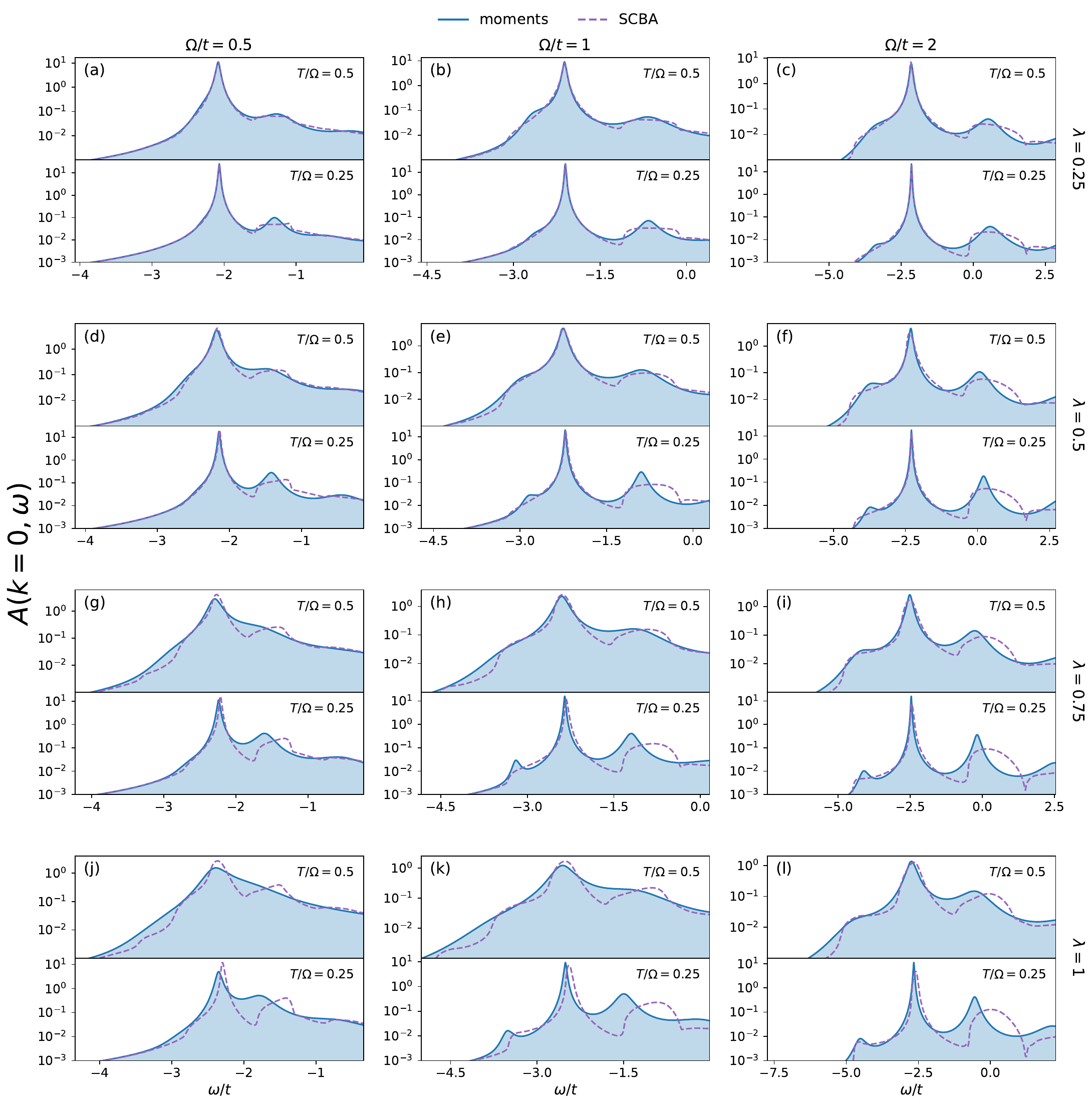}
\caption{\label{fig:ssh_thermal_log}The SSH spectra of Fig.~\ref{fig:ssh_thermal} on separate logarithmic axes for the two temperatures, $T/\Omega=0.25$ below and 0.5 above. t.}
\end{figure*}
\subsection{Holstein and SSH spectra from moments}
\label{sec:spectra}

Using the techniques and intuition developed in the past sections, we apply the reconstruction method to the Holstein polaron, whose interaction changes the energy of an occupied site, and the site-Peierls or SSH polaron, whose interaction moves the electron while creating or annihilating a phonon. Both have $\varepsilon_k=-2t\cos k$ and Einstein phonons. We use the couplings of Table~\ref{tab:vertices}: $\lambda=g^{2}/(2t\Omega)$ for Holstein and $\lambda=2g^{2}/(t\Omega)$ for SSH, with $g$ in the latter case the real-space bond-modulation amplitude. The spectra in this subsection are at $k=0$; this is the ground-state momentum of the Holstein polaron, but not necessarily that of the SSH polaron - the ground state momentum shifts at a critical coupling value \cite{Marchand2010, Marchand2017}.

The input is $M_0$--$M_{28}$ for Holstein and $M_0$--$M_{23}$ for SSH. The former fixes $a_0,\ldots,a_{13}$ and $b_1,\ldots,b_{14}$, while the latter fixes $a_0,\ldots,a_{11}$ and $b_1,\ldots,b_{11}$. At $T=0$ both $a_n$ and $b_n$ are continued linearly, with slopes fitted to the last four exact values. At $T>0$ we hold $a_n$ constant and continue only $b_n$ linearly. 

For Holstein we compare with MA$^{(1)}$, including its finite-temperature extension~\cite{Shannigrahi2026}, and with SCBA. For SSH the zero-temperature comparison is with three-site MA$^{(0)}$ and SCBA; the positive-temperature comparison contains only SCBA, because the SSH MA calculation used here is a zero-temperature one. The three-site approximation allows the phonons to occupy three consecutive sites while the electron can move away from the cloud~(details can be found in Ref.\cite{BerciuFehske2010,Marchand2017}). Data from multi-site MA$^{(0)}$ and MA$^{(1)}$ for the finite-T SSH problem is reserved for future work.  In the absence of heavy numerically exact data, we treat the MA results as closest to exact results. Accuracy of MA results is well established in the literature.

 In all figures the moment reconstruction is blue (with filled in light blue), MA is orange (dash-dotted), and SCBA is purple (dashed).

Fig.~\ref{fig:spectra} show the Holstein spectra for $\Omega/t=0.5$, 1, 2 and $\lambda=0.1$, 0.5, 1.0, with $\eta=0.05t$ for three representative temperatures ($T/\Omega=0.25$ and 0.5). At weak coupling the three methods give nearly coincident quasiparticle peaks at all three temperatures. As expected, increasing $\lambda$ separates the SCBA peak since it becomes increasingly inaccurate. Interestingly,  the lower-energy peaks of the moment reconstruction and MA$^{(1)}$ overlap to a great degree. In the bottom-most band, the separation is particularly clear in panels (g)--(i), whereas the moment and MA$^{(1)}$ peaks remain much closer.   This visual comparison is consistent with the energy-bound test of Sec.~\ref{sec:bounds_vs_methods}, which does not use a terminator. 

At higher temperatures, the contrast between the methods becomes clear with higher coupling values. At $\lambda=0.5$ the peak height decreases with temperature and the sideband structure becomes less distinct, while the differences in peak positions and heights remain visible. The $\lambda=1$ row gives the clearest limitation: the moment reconstruction from the constant-$a$/linear-$b$ terminator produces a broad continuum where thermal MA$^{(1)}$ retains comparatively sharp structure. This occurs most strongly at $\Omega/t=0.5$ and 1 and becomes more pronounced at $T/\Omega=0.5$. Throughout, the moment and MA$^{(1)}$ spectra remain much closer than SCBA near the qp peak.

The logarithmic view in Fig.~\ref{fig:hp_spectra_log} allows one to see while there is quantitative agreement at the qp peak between MA$^{(1)}$ and the reconstructed spectra, there are appreciable differences in the placement of the polaron+one phonon and higher order continua (as explained in earlier sections). Instead of being a true continuum, it appears as a series of peaks in the  reconstructed spectra.

Figures~\ref{fig:ssh_spectra} and \ref{fig:ssh_spectra_log} extend the comparison to SSH at $T=0$, using $\eta=0.02t$ and a wider coupling range, $\lambda=0.1$, 0.5, 1.0, 2.0, 3.0. At $\lambda=0.1$ the qp peak again agrees closely among all three methods; the logarithmic scale reveals differences in the continuum. By $\lambda=1$, the moment reconstruction and three-site MA$^{(0)}$ retain nearby low-energy peaks and more structured higher-energy weight, whereas SCBA places the qp peak at higher energy. At $\lambda=2$ and 3, SCBA retains a large qp peak where the other methods distribute weight among several features. 

We should note a qualitative difference from Holstein. Beyond the SSH momentum transition, $k=0$ no longer labels the ground state. The diminishing low-energy weight in the $k=0$ spectra, especially in the strong-coupling adiabatic panels, may not be the ground-state.

The SSH thermal comparison, Figs.~\ref{fig:ssh_thermal} and \ref{fig:ssh_thermal_log}, uses $\lambda=0.25$, 0.5, 0.75, 1, $\Omega/t=0.5$, 1, 2, and $\eta=0.01t$. At $\lambda=0.25$ moments and SCBA give very similar main peaks at both temperatures, although their continuum description differs on the logarithmic scale. With increasing coupling, the moment peak lies lower in energy than the SCBA peak, most visibly for $\Omega/t=0.5$ and 1. Thermal continuum below the qp peak are especially clear for $\Omega/t=2$, where their separation is larger.

\begin{figure*}[tp]
\centering
\includegraphics[width=0.49\textwidth]{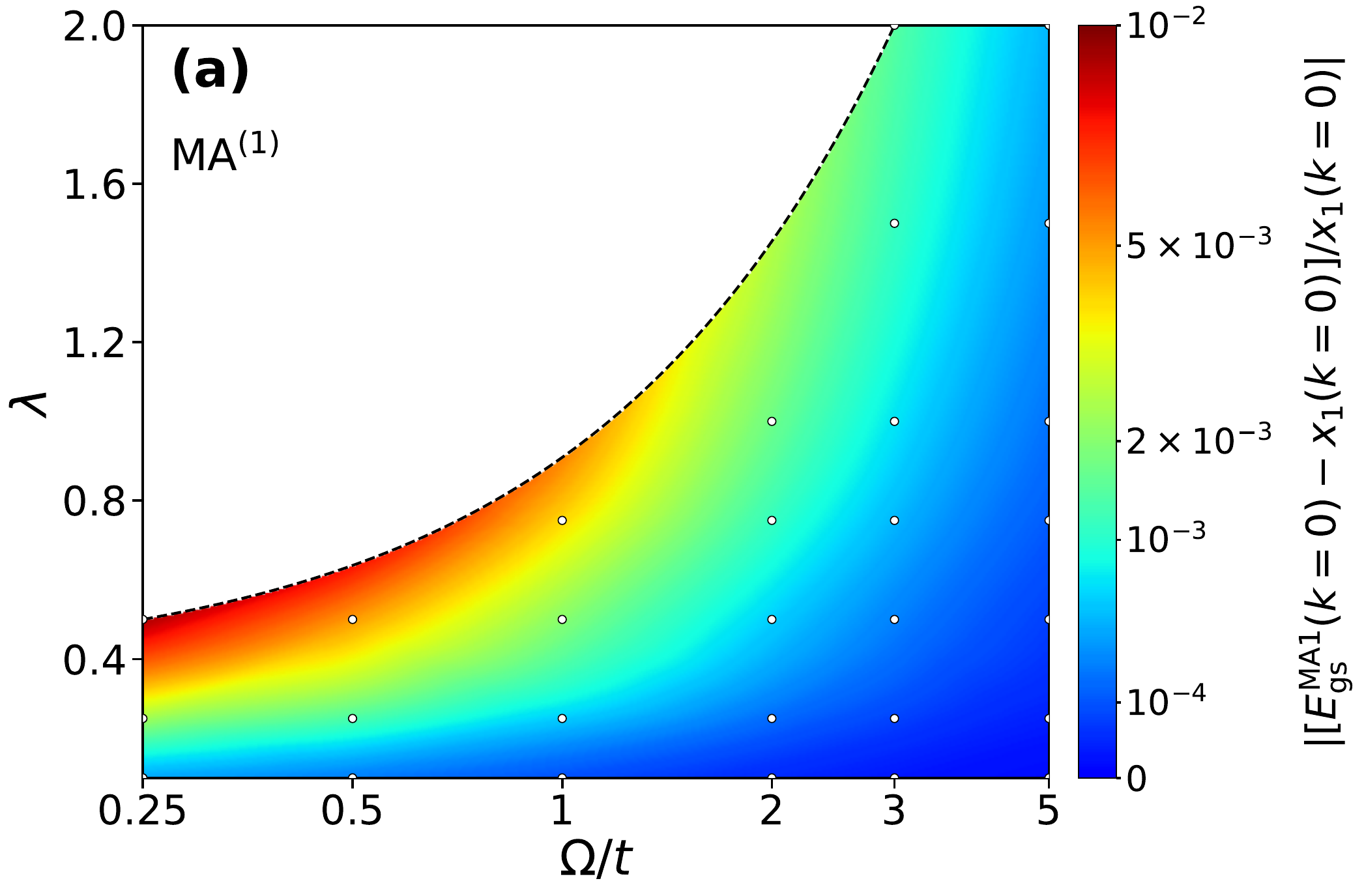}\hfill
\includegraphics[width=0.49\textwidth]{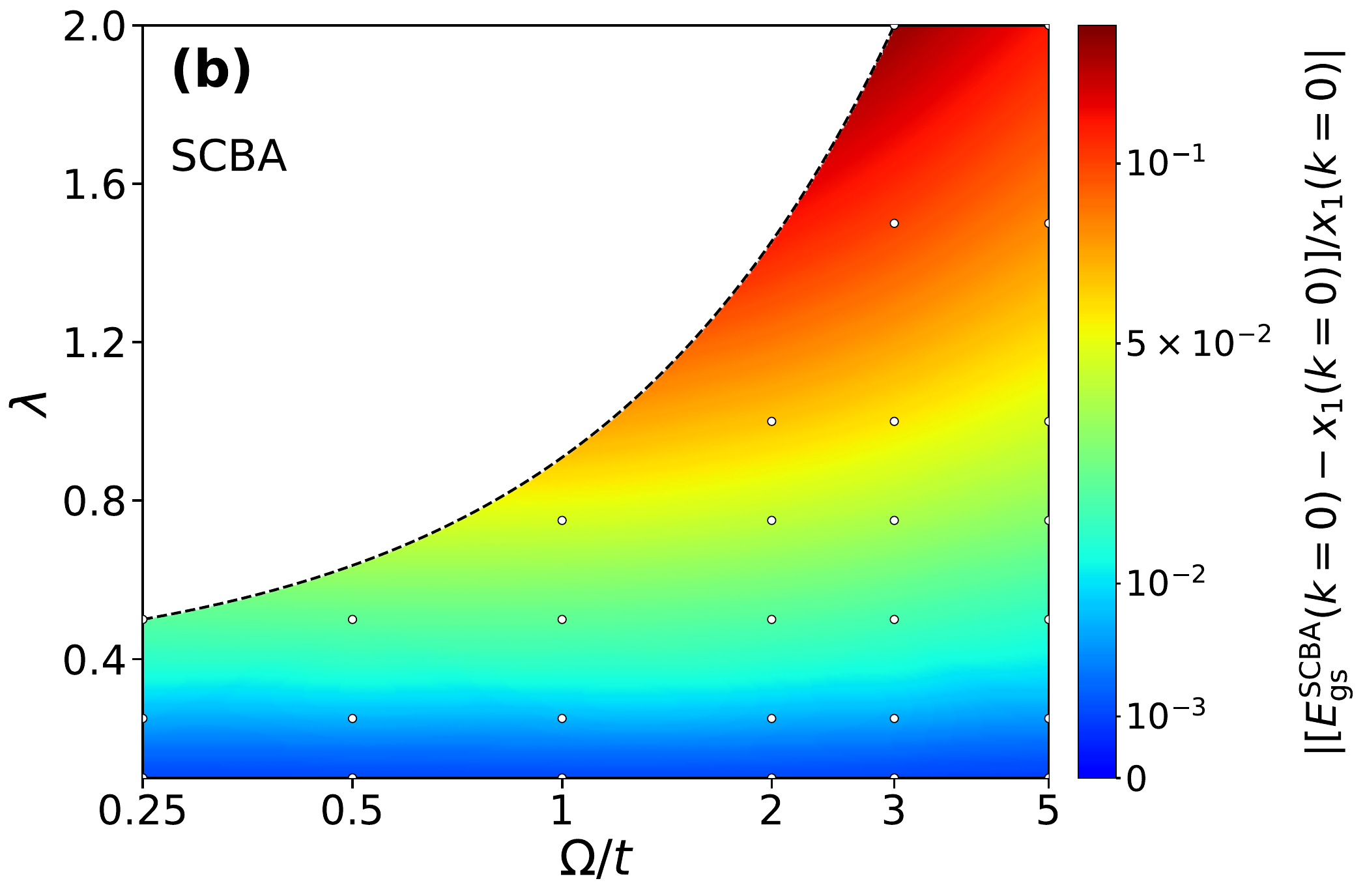}
\caption{\label{fig:bounds_heat}Holstein ground-state relative energy differences$\left|\frac{E_{\rm gs}^{\rm *}(k=0)-x_1^{(14)}(k=0)}{x_1^{(14)}(k=0)}\right|$ for (a) MA$^{(1)}$ and (b) SCBA. The lowest Gauss node is built from the exact moments and has no broadening; method energies are peak fits at $\eta=0.01t$. Circles mark displayed computed points. The display omits the region above the dashed boundary joining $(\Omega/t,\lambda)=(0.25,0.5)$ to $(3,2)$, where the bound convergence is not sufficient. The two colour scales differ substantially.}
\end{figure*}

\section{Applications of the moments}
\label{sec:applications}

The spectral comparisons contain an unavoidable choice of tail. The finite Jacobi matrix does not depend on the tail and provide rigorous upper bounds on zero-temperature energies. We first formulate these at fixed momentum and for the global ground state, use them to compare Holstein and SSH approximations, and then return to the reconstructed SSH dispersion to examine its momentum transition. The bounds and the reconstructed dispersion are different uses of the same exact input and must be interpreted differently.

\begin{figure*}[tp]
\centering
\includegraphics[width=0.49\textwidth]{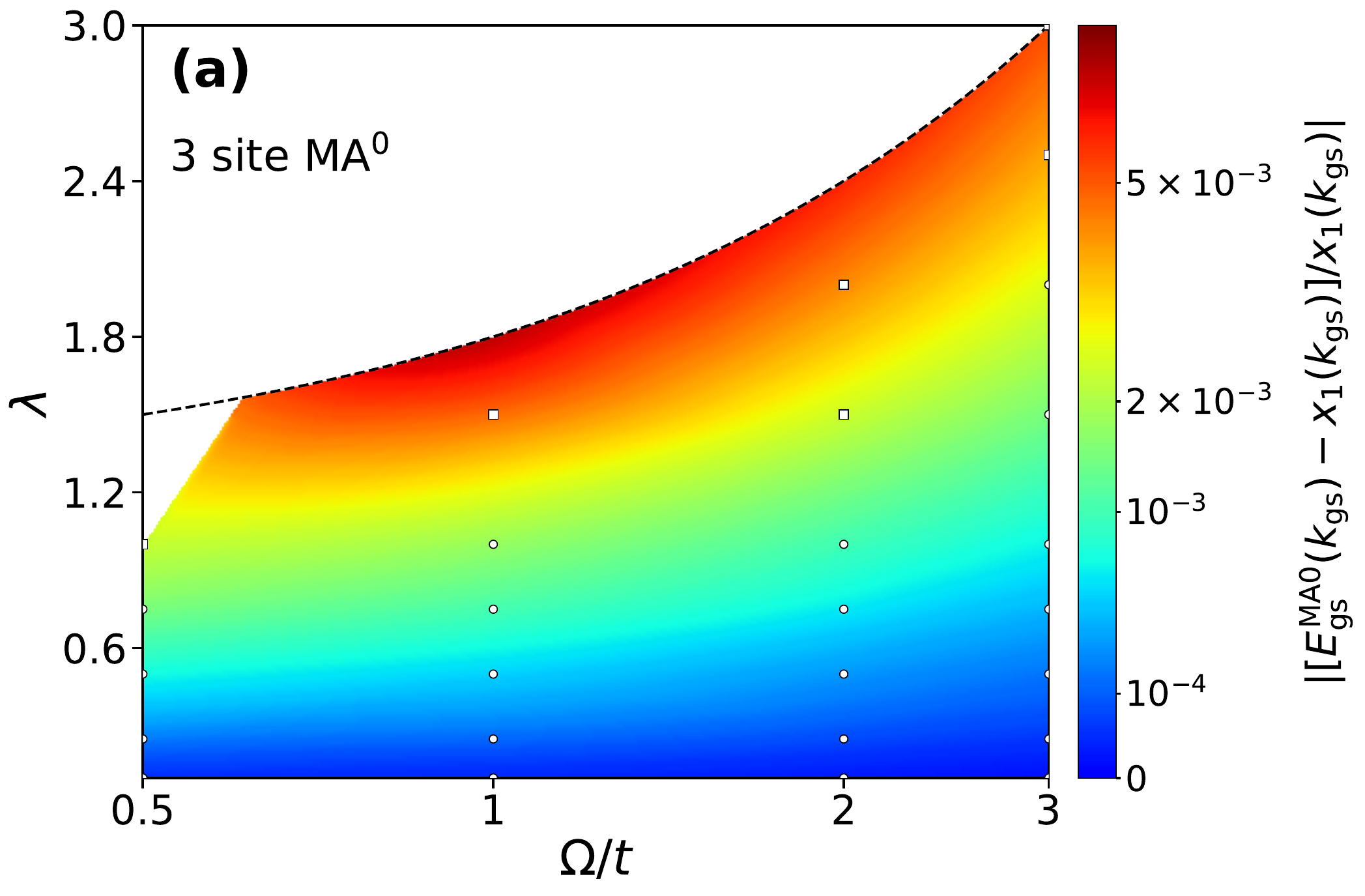}\hfill
\includegraphics[width=0.49\textwidth]{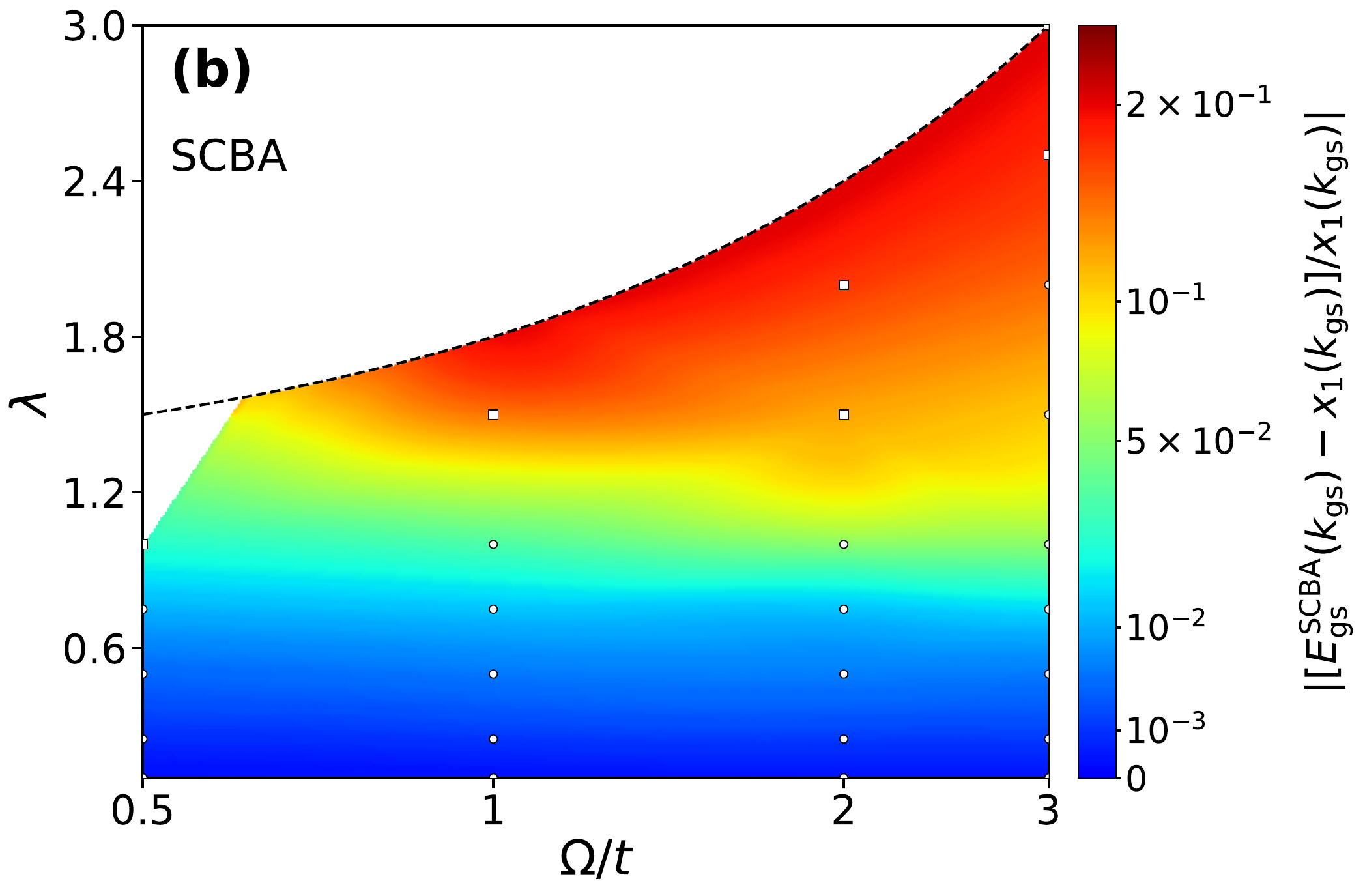}
\caption{\label{fig:ssh_bounds_heat}SSH ground-state energy differences $\left|\left[E_{\rm gs}^{*}(k_{\rm gs})-x_1^{(12)}(k_{\rm gs})\right]/x_1^{(12)}(k_{\rm gs})\right|$ for (a) three-site MA$^{(0)}$ and (b) SCBA, with $U_{12}=\min_k x_1^{(12)}(k)$ from $M_0$--$M_{23}$. The two colour scales differ substantially.}
\end{figure*}

\subsection{Rigorous bounds at $T=0$}
\label{sec:bounds}

At each momentum, construct $J_N(k)$ from the moments in that momentum sector. Let $x_1^{(N)}(k)$ be its lowest eigenvalue. Projection onto the nested Krylov spaces of Sec.~\ref{sec:CF} gives
\begin{equation}
E_{\rm sector}(k)
\le x_1^{(N+1)}(k)\le x_1^{(N)}(k),
\label{eq:sector_bound}
\end{equation}
where $E_{\rm sector}(k)$ is the lowest energy in the full momentum sector. The Rayleigh--Ritz principle gives the upper bound\cite{HornJohnson2013}. We can trivially understand why there is a monotonic improvement with $N$, enlarging the variational space can only lower the estimate of the ground state energy.

For Holstein the ground state occurs at $k=0$. For SSH one must minimize over momentum:
\begin{equation}
E_{\rm gs}\le U_N\equiv\min_{k\in\mathrm{BZ}}x_1^{(N)}(k),
\qquad U_{N+1}\le U_N.
\label{eq:global_bound}
\end{equation}
A separate Jacobi matrix is needed at every $k$;  We do a search on $0\le k\le\pi/2$.  These bounds are a function of the exact set of finite moments and do not depend on extrapolated coefficients or artificial spectral broadening.

The lowest Gauss node converges rapidly with order in $N$ and more exact moments tighten the bound. For the Holstein case, $N=14$ allows us to converge well beyond $10^4$ for a vast majority of the parameter space. Specifically for the Holstein example $\Omega=t$, $\lambda=0.5$, the node sequence gives $x_1^{(13)}\simeq-2.469684t$, very close to the variational exact-diagonalization value $-2.469684724t$~\cite{Bonca1999}

There is also a weight inequality at fixed momentum. With Gauss nodes $x_j$ and weights $w_j$, the polynomial $q(\omega)=\prod_{j=2}^{N}(\omega-x_j)/(x_1-x_j)$ obeys $q(x_j)=\delta_{1j}$ and $q(\omega)^2\ge1$ for $\omega\le x_1$. This imply
\begin{equation}
\int_{-\infty}^{x_1}A(k,\omega)\,d\omega
\le\int q(\omega)^2 A(k,\omega)\,d\omega=w_1.
\label{eq:weight_bound}
\end{equation}
Thus $Z(k)\le w_1^{(N)}(k)$ for a lowest isolated pole. At $T>0$, it is no longer useful as there is spectral weight below the qp ground state from thermal continua.

\subsection{Limits of SCBA and MA approximations through moment-derived bounds}
\label{sec:bounds_vs_methods}

Let an approximation predict a ground-state energy $E_{\rm gs}^{\rm *}$, minimized over its own momentum when necessary. Its distance above the moment bound is
\begin{equation}
D_N^{\rm app}=E_{\rm gs}^{\rm *}-U_N,
\qquad E_{\rm gs}^{\rm *}-E_{\rm gs}\ge D_N^{\rm app}.
\label{eq:method_distance}
\end{equation}
Every positive $D_N^{\rm app}$ therefore supplies a lower bound on the approximation's energy error. The test is independent of the spectral tail.

\paragraph*{Holstein.} Figure~\ref{fig:bounds_heat} uses the $k=0$ states over $\Omega/t=0.25$--5 and $\lambda=0.1$--2. The MA$^{(1)}$ and SCBA energies were extracted from their lowest peaks at $\eta=0.01t$ by a Lorentzian fit. The white region omits the strong-coupling adiabatic region where the convergence of the bounds is not below$10^{-4}$. 

SCBA lies above the bound at every displayed point, considerably so (nearly an order of magnitude higher than MA$^{(1)}$). The corresponding MA$^{(1)}$ distances are within the tolerance of convergence and is the worst ($\approx 1\%$) in the strongly adiabatic regime.

\paragraph*{SSH.} Figure~\ref{fig:ssh_bounds_heat} performs the analogous comparison with the global bound $U_{12}$. Both MA$^{(0)}$ and SCBA are minimized over their own momentum, independently of the momentum that minimizes the Gauss node. The data contains 34 parameter cases: $\Omega/t=0.5$-3  with $\lambda=0.1-3$. The observations remain similar to the Holstein case, with arguably even less deviation of the  MA$^{(0)}$ data.

\section{Summary and outlook}
\label{sec:conclusions}

The frequency moments of the single-electron addition spectrum provide exact, closed-form constraints on an electron--phonon problem without requiring its full solution. We have solved the problem of calculating closed form expressions for higher order moments by replacing the direct expansion of nested commutators by a recurrence for phonon-number-resolved amplitudes. At $T=0$ it has three terms; TFD converts the problem into one with two boson species and gives a five-term recurrence at $T>0$. The formulation is agnostic to the electron dispersion, phonon dispersion, and coupling vertex and takes them as an adjustable input. We specifically show the results for Holstein, site-Peierls/SSH models with coupling to Einstein phnons.

We establish the relevant applications of these moments. As exact sum rules they identify where a technique's approximate Green's function first drops a term in the diagrammatic expansion. In parallel, through the Jacobi continued fraction formulation, they allow reconstruction of the polaron spectra. The Holstein and SSH comparisons illustrate both the usefulness and the limits of the latter construction. At weak coupling the qp agree closely with close-to-exact data from MA, including at finite T. At larger coupling, while the moment spectra differs from MA, it does so substantially less than SCBA.

Furthermore, the finite Jacobi matrix yields a rigorous upper bound on the polaron ground-state energy. This allows us to compute the qp energy from existing methods and identify their deviation. SCBA deviates to a large degree through a large portion of the parameter space while MA discrepancies are much smaller and limited to the strongly adiabatic regime. 

Our analysis extends possible applications of moments besides being used as a trivial check on the sanity of numerical/experimental spectral data. While less accurate than MA, this model-agnostic approach opens the door to unexplored or numerically intractable models, before computationally demanding exact methods are applied. 
\\

\begin{acknowledgments}
J.S.\ thanks P.~E.~Kornilovitch and P.~Mitri\'c for useful discussions. J.S highlights the role of Sec IIIB in Ref.\cite{Goodvin2006} and Mr. Ovidiu Toader's initial Mathematica conception in 2005-2006 for calculating moments for the Holstein polaron which partly inspired this whole work.  This project was supported in part by the Max Planck--UBC--UTokyo Center for Quantum Materials. J.S.\ acknowledges funding from the Natural Sciences and Engineering Research Council of Canada in Quantum Computing Program, grant number 543245, the C.D. Howe Graduate Fellowship from the University of British Columbia and from Mitacs Globalink Research Award IT46276. M.B.\ acknowledges funding from the Natural Sciences and Engineering Research Council of Canada.
\end{acknowledgments}

\bibliographystyle{apsrev4-2}
\bibliography{sum_rules_refs}

\end{document}